\documentclass[aps,prc,reprint,amsmath,amssymb,nofootinbib,longbibliography,floatfix]{revtex4-2}

\usepackage{graphicx}
\usepackage{booktabs}
\usepackage{tikz}
\usetikzlibrary{arrows.meta,positioning}

\graphicspath{{plots/}}

\newcommand{\kf}{k_{\mathrm{F}}}
\newcommand{\GeV}{\,\mathrm{GeV}}
\newcommand{\MeV}{\,\mathrm{MeV}}
\newcommand{\code}[1]{\texttt{#1}}
\newcommand{\si}[1]{\ensuremath{\mathrm{#1}}}

\begin{document}

\title{LUNAR: a Monte Carlo generator for bound-nucleon decay in liquid argon}

\author{J.\,A.\ Nowak}
\email[Corresponding author: ]{j.nowak@lancaster.ac.uk}
\affiliation{School of Physics and Astronomy, Lancaster University,
Lancaster LA1 4YB, United Kingdom}

\date{\today}

\begin{abstract}
The search for nucleon decay in liquid-argon time-projection chambers requires a
quantitative description of how the bound nuclear environment reshapes the
decay-product kinematics. We present LUNAR, a fast, openly available Monte Carlo generator dedicated to two-body decays of protons and neutrons bound in
argon-40, the target nucleus of the DUNE far detector. The parent nucleon is
drawn from a selectable nuclear ground state (ten momentum distributions
ranging from mean-field Fermi gases to argon spectral functions) and
bound off the mass shell by one of three removal-energy prescriptions, including
the momentum-dependent optical potential of
Juszczak \textit{et al}. The two-body decay is performed off-shell and
boosted to the laboratory frame, and the daughter meson is then propagated out
of the nucleus by a semi-classical intranuclear cascade with an optional
formation zone for the freshly produced meson. We use the generator to
separate the distinct roles of Fermi motion and binding in shaping the observable
meson spectrum, to quantify final-state interactions channel by channel, and to
translate present Super-Kamiokande limits into expected DUNE event yields for the
full set of standard decay modes. Final-state interactions leave the
supersymmetry-favored $p\to K^{+}\bar\nu$ signal, and the neutral-kaon modes,
essentially intact while roughly
halving the pion, $\eta$, and antikaon rates, an effect that dominates over the
$\sim$10\% spread induced by the choice of nuclear model. For the pion channels
the modeling uncertainty of the cascade itself exceeds that spread. The code is released to
the community as a lightweight, extensible tool for signal efficiency and
systematics studies.
\end{abstract}

\maketitle

\section{Introduction}
\label{sec:intro}

The conservation of baryon number is an accidental symmetry of the Standard
Model, not supported by any gauge principle, and essentially every attempt to
embed the Standard Model into a more unified framework predicts that it is
broken. Grand unified theories are the canonical example: by placing quarks and
leptons in common multiplets they introduce interactions that convert one into
the other, so that the proton (and the neutron bound inside a nucleus) becomes
unstable. The minimal $SU(5)$ theory of Georgi and
Glashow~\cite{GeorgiGlashow1974}, rooted in the quark--lepton unification of
Pati and Salam~\cite{PatiSalam1974}, mediates decay through dimension-six
operators~\cite{Weinberg1979,WilczekZee1979} and singles out
$p\to e^{+}\pi^{0}$ as the leading mode. Supersymmetric grand unification changes
this picture qualitatively: dimension-five operators generated by the exchange of
colored Higgsinos dominate, and because they favor a strange quark in the final
state the preferred channel becomes $p\to K^{+}\bar\nu$~%
\cite{SakaiYanagida1982,DimopoulosRabyWilczek1982}. Which mode is dominant is
therefore a direct diagnostic of the underlying unification scheme, and a broad
program of searches across many channels is needed to discriminate between
them~\cite{Nath2007,Babu2013}. Baryon-number violation need not stop at single
nucleon decay; dinucleon transitions and neutron--antineutron oscillation probe
complementary operators~\cite{PhillipsNNbar2016}, and a flexible simulation tool
should anticipate them.

Liquid-argon time-projection chambers, and in particular the far detector of the
Deep Underground Neutrino Experiment (DUNE), are exceptionally well matched to
the supersymmetry-favored channel~\cite{DUNE2020}. A kaon produced in the decay
of a free proton at rest would be monochromatic and, at the relevant momentum,
would stop and decay inside the active volume, leaving a short, heavily ionizing
track with a distinctive decay topology. This combination of calorimetric and
tracking information makes $p\to K^{+}\bar\nu$ the golden liquid-argon channel.
The same detector retains sensitivity to the full complement of standard
two-body modes: $e^{+}\pi^{0}$, $\mu^{+}\pi^{0}$, $\bar\nu\pi$, the $e^{+}$ and
$\mu^{+}$ companions of the $\eta$, and the neutral- and charged-kaon
channels, each of which already carries a dedicated experimental limit
(Sec.~\ref{sec:channels}).

The free-nucleon case, however, is never realized in a detector. The decaying
nucleon is bound, so it carries  momentum and sits below its free mass by a
removal energy that depends on the nuclear shell it occupies. Both effects act on
the observable: Fermi motion smears the otherwise sharp daughter spectrum, while
binding pushes the decaying system off the mass shell, lowering the available
energy and, for the heavier final states, forbidding a fraction of decays
outright. The decay meson must then traverse the residual nucleus, where it can
scatter, charge-exchange, or be absorbed before it ever reaches the detector. An
honest estimate of the signal efficiency, of the reconstructed meson spectrum,
and of the accompanying hadronic activity therefore demands a controlled model of
the nuclear initial state and of the meson's passage out of the nucleus.

These nuclear effects are, in principle, within reach of the general-purpose
neutrino event generators GENIE~\cite{AndreopoulosGENIE2010},
NuWro~\cite{NuWro2025}, and GiBUU~\cite{BussGiBUU2012}, whose nuclear models and
intranuclear cascades were built for the same argon target. Those frameworks,
however, are engineered around the neutrino--nucleus interaction vertex; using
them to study nucleon decay means working against their intended design, and
their breadth comes at the cost of weight and opacity. Recent work has begun to
quantify these effects specifically for proton decay: a GiBUU transport
calculation~\cite{Yan2026} for $p\to e^{+}\pi^{0}$ in water Cherenkov detectors,
and a survey by Barbu \textit{et al.}~\cite{Barbu2025} of Fermi motion and
final-state interactions for $p\to\pi^{+}\bar\nu$ and $p\to K^{+}\bar\nu$ across
liquid argon, xenon, and water. Both, however, either repurpose a heavy
general-purpose transport code or work outside an event-generator framework
altogether. There is consequently a
gap for a small, transparent, dedicated tool, in the spirit of compact
single-purpose generators such as MARLEY~\cite{GardinerMARLEY2021}, that an
analyzer can read in an afternoon, run in seconds, and re-purpose to a new
channel or nuclear model by editing a single header. LUNAR (the Lancaster
University Nucleon-decay ARgon generator) is built to fill that gap. It samples a
bound nucleon from a menu of nuclear models, performs the off-shell two-body
decay, transports the daughter meson through a semi-classical cascade, and writes
a complete final-state record, for any of the standard proton and neutron
channels.

In this paper we describe the physics content of the generator
(Sec.~\ref{sec:generator}) and use it to address three questions of direct
relevance to a liquid-argon nucleon-decay search. First, we disentangle the
separate effects of Fermi motion and nuclear binding on the observable meson
spectrum, showing that the momentum model alone fixes its \emph{width} while both
ingredients move its \emph{central value}, the momentum model by the larger
amount (Sec.~\ref{sec:results}). Second, we
quantify final-state interactions channel by channel and demonstrate that they
are negligible for the strange modes but a leading, strongly mode-dependent
correction for the pion, $\eta$, and antikaon channels. Third, we fold these
ingredients into expected DUNE event yields for every standard mode, anchored to
current Super-Kamiokande limits. We then discuss how the generator can be used
and extended by the community (Secs.~\ref{sec:usage} and~\ref{sec:extensions})
before concluding (Sec.~\ref{sec:summary}).

\section{The generator}
\label{sec:generator}

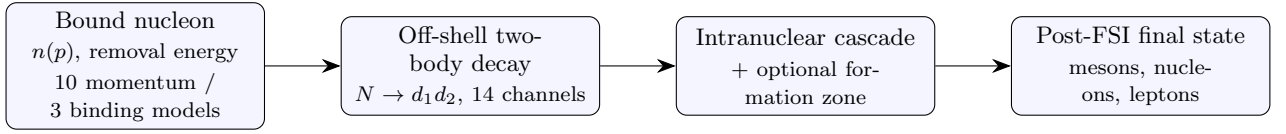
\begin{figure*}[t]
  \centering
  \begin{tikzpicture}[
      box/.style={draw, rounded corners, align=center, minimum height=13mm,
                  text width=32mm, font=\small, fill=blue!4},
      >={Stealth[length=2.5mm]}]
    \node[box] (gs) {Bound nucleon\\[1pt]
      {\footnotesize $n(p)$, removal energy}\\[1pt]
      {\footnotesize 10 momentum / 3 binding models}};
    \node[box, right=10mm of gs] (decay) {Off-shell two-body decay\\[1pt]
      {\footnotesize $N\to d_1 d_2$, 14 channels}};
    \node[box, right=10mm of decay] (cas) {Intranuclear cascade\\[1pt]
      {\footnotesize $+$ optional formation zone}};
    \node[box, right=10mm of cas] (fin) {Post-FSI final state\\[1pt]
      {\footnotesize mesons, nucleons, leptons}};
    \draw[->] (gs) -- (decay);
    \draw[->] (decay) -- (cas);
    \draw[->] (cas) -- (fin);
  \end{tikzpicture}
  \caption{The LUNAR generation pipeline. A bound nucleon is drawn from the
  chosen momentum and binding model, decayed off the mass shell into the selected
  channel, and its hadron is transported out of the nucleus by the intranuclear
  cascade (with an optional formation zone), yielding the complete post-FSI final
  state. Each stage is an independent, replaceable module.}
  \label{fig:pipeline}
\end{figure*}

LUNAR is written as a header-only \code{C++17} physics library driven by a small
set of command-line executables. It has no external dependencies beyond the
standard library; the ROOT framework is used only by the downstream analysis
macros, not by the generator itself. A single run samples the bound-nucleon
ground state, performs the off-shell decay, propagates the products through the
nucleus, and streams the final state to a plain-text file
(Fig.~\ref{fig:pipeline}). The physics is
organized into independent modules: nuclear momentum and binding, off-shell
kinematics, the spectral functions, the hadron--nucleon cross sections,
and the cascade. Each of these can be modified or replaced without touching the
others. This modularity is deliberate: it is what makes the extensions of
Sec.~\ref{sec:extensions} a matter of local edits rather than redesign.

\subsection{Initial-state nucleon momentum}
\label{sec:momentum}

The argon-40 ground state is described in the local-density approximation. The
nucleon density follows a two-parameter Fermi (Woods--Saxon) form~%
\cite{Juszczak2005},
\begin{equation}
  \rho(r) = \frac{\rho_0}{1 + \exp\!\big[(r-C)/C_1\big]},
  \qquad
  \rho_0 = 0.176\ \mathrm{fm^{-3}},
  \label{eq:density}
\end{equation}
with $C = 3.530\ \mathrm{fm}$ and $C_1 = 0.541\ \mathrm{fm}$. Treating each
nucleon species as a locally cold Fermi gas, the position-dependent Fermi
momentum is
\begin{equation}
  \kf^{q}(r) = \hbar c\,\Big(3\pi^2\,\tfrac{N_q}{A}\,\rho(r)\Big)^{1/3},
  \label{eq:localkf}
\end{equation}
evaluated separately for protons and neutrons with the argon isospin fractions
$Z/A = 0.45$ and $N/A = 0.55$. For protons the density-weighted average
reproduces the reference value $\kf^{p} = 217\MeV/c$, which also
serves as the fixed Fermi momentum of the global-gas option; the corresponding
neutron value is $\kf^{n} = 230\MeV/c$ (Appendix~\ref{app:models}). Unless stated
otherwise the numbers quoted below refer to proton modes, and $\kf$ means
$\kf^{p}$.

The user selects the nucleon momentum distribution $n(p)$ from the ten
physics-based models of Table~\ref{tab:models}; an eleventh option, a polynomial
parametrization (\code{poly}) whose coefficients are read from a configuration
file, is a deliberately simple toy that can be tuned to prototype an arbitrary
$n(p)$ for testing rather than to represent a specific nuclear model. The ten
physics models span a deliberate range of physical assumptions.
The global Fermi gas is a hard sphere with a sharp surface at $\kf$ and no
strength above it. The local-density variants soften that surface, populating low
momenta because a decay can occur in the dilute nuclear periphery. The
correlated models (short-range correlations, the Bodek--Ritchie
gas~\cite{BodekRitchie1981}, and the correlated Fermi gas) restore the
high-momentum nucleons that a pure mean field omits, adding a $1/p^4$ tail above
$\kf$. The harmonic-oscillator shell model replaces the sphere altogether by a
sum of single-particle orbitals, producing genuine shell structure, while the
Gaussian provides a smooth reference with no Fermi surface at all. Finally, the
two spectral-function options supply their own joint momentum and removal-energy
distribution (Sec.~\ref{sec:spectral}): \code{benhar} draws the nucleon directly
from a tabulated argon spectral function, while \code{ankowski} builds an
effective spectral function analytically from the argon shell structure and a
correlated tail.

\begin{table}[!b]
  \centering
  \footnotesize
  \setlength{\tabcolsep}{4pt}
  \begin{tabular}{@{}lll@{}}
    \toprule
    Key & Model & $n(p)$ \\
    \midrule
    \code{poly} & Polynomial (toy) & $f(p)$ from config file \\
    \code{gfg}  & Global Fermi gas & $\propto p^2$, $p<\kf$ \\
    \code{lfg}  & Local Fermi gas  & sphere $\kf(r)$, $r^2\rho(r)$-weighted \\
    \code{src}  & Short-range corr. & Fermi bulk $+$ $1/p^4$ tail \\
    \code{sf}   & Spectral fn.\ (toy) & local Fermi gas $+$ tail \\
    \code{hosm} & HO shell model & $\sum_{nl}\mathrm{occ}\,|\phi_{nl}(p)|^2$ \\
    \code{br}   & Bodek--Ritchie & Fermi bulk $+$ long $1/p^4$ tail \\
    \code{gauss} & Gaussian & $\propto p^2\exp(-p^2/2\sigma^2)$ \\
    \code{cfg}  & Correlated FG & sharp sphere $+$ contact tail \\
    \code{benhar} & Tabulated $S(p,E)$ & NuWro argon grid \\
    \code{ankowski} & Analytic $S(p,E)$ & shell mean field $+$ $1/p^4$ tail \\
    \bottomrule
  \end{tabular}
  \caption{Initial-state nucleon momentum models available in LUNAR. Parameter
  choices (high-momentum fractions, cut-offs, oscillator length) are summarized
  in Appendix~\ref{app:models}.}
  \label{tab:models}
\end{table}

\begin{figure}[tbp]
  \centering
  \includegraphics[width=\columnwidth]{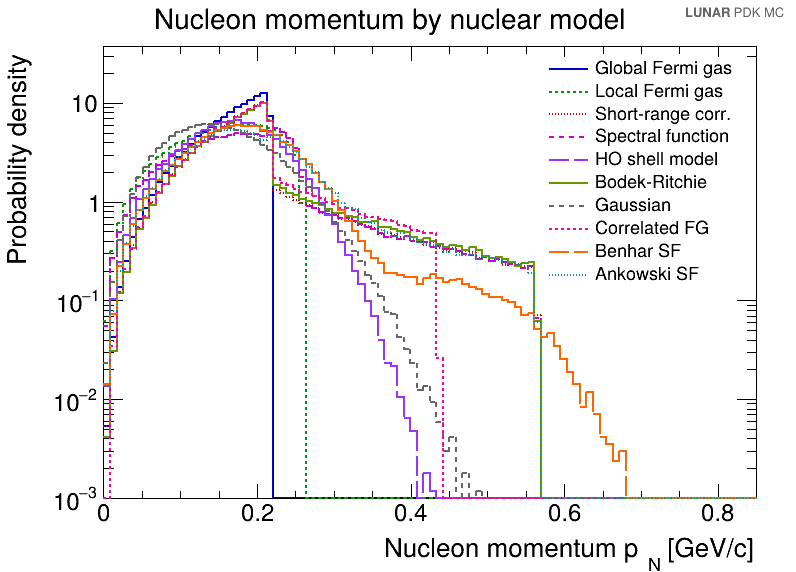}
  \caption{Proton momentum distributions generated by LUNAR for the nuclear
  models of Table~\ref{tab:models} (area-normalized, logarithmic vertical scale).
  The local-density treatment smears the Fermi-gas edge near $\kf$, while the
  short-range-correlation, Bodek--Ritchie, and correlated-Fermi-gas options
  extend well beyond it; the Gaussian falls off smoothly and the
  harmonic-oscillator shell model shows orbital structure in place of a sharp
  surface. Models are distinguished by line style as well as by color.}
  \label{fig:models}
\end{figure}

The resulting distributions are overlaid in Fig.~\ref{fig:models}, and a few
 observables are collected in Table~\ref{tab:momsummary}: the mean and
root-mean-square of the nucleon momentum, the position of the distribution peak,
and the fraction of nucleons carrying more than $\kf$. The latter is the cleanest
discriminator between the model families. The global Fermi gas, by construction,
has none; the local-density and shell models put a tenth to a fifth of the
strength above $\kf$; and the correlated models and the spectral
functions carry between a fifth and a third there. As we show in
Sec.~\ref{sec:results}, this high-momentum content is what controls the width of
the observable daughter spectrum.

\begin{table}[!t]
  \centering
  \small
  \begin{tabular}{@{}lrrrr@{}}
    \toprule
    Model & $\langle p_N\rangle$ & RMS & peak & $p_N>\kf$ \\
          & [GeV/c] & [GeV/c] & [GeV/c] & [\%] \\
    \midrule
    Global Fermi gas    & 0.163 & 0.042 & 0.208 & 0.0 \\
    Local Fermi gas     & 0.157 & 0.058 & 0.174 & 17.5 \\
    Short-range corr.   & 0.196 & 0.088 & 0.208 & 18.8 \\
    Spectral function   & 0.191 & 0.097 & 0.225 & 32.9 \\
    HO shell model      & 0.168 & 0.060 & 0.174 & 20.8 \\
    Bodek--Ritchie      & 0.199 & 0.090 & 0.208 & 20.7 \\
    Gaussian            & 0.155 & 0.065 & 0.149 & 17.1 \\
    Correlated FG       & 0.190 & 0.072 & 0.208 & 19.9 \\
    Benhar SF           & 0.189 & 0.080 & 0.174 & 30.4 \\
    Ankowski SF         & 0.200 & 0.094 & 0.174 & 35.4 \\
    \bottomrule
  \end{tabular}
  \caption{Initial nucleon-momentum observables by model
  ($2\times10^5$ $p\to K^{+}\bar\nu$ events, optical-potential binding). The
  high-momentum fraction $p_N>\kf^{p}=0.217\GeV/c$ rises from zero for the
  hard-edged global gas to roughly a third for the correlated and
  spectral-function models. The peak is the center of the most populated bin of
  Fig.~\ref{fig:models}.}
  \label{tab:momsummary}
\end{table}

\subsection{Binding and off-shell kinematics}
\label{sec:binding}

A bound nucleon is not a free particle at rest: it is off the mass shell by an
amount governed by the nuclear potential it feels. LUNAR offers three
interchangeable prescriptions for the removal energy of a mean-field nucleon,
selected at run time with the \code{--binding} flag:
\begin{itemize}
  \item \code{potential}: the momentum-dependent optical potential of
        Eq.~\eqref{eq:potential} below (the default), which is the deepest and
        the only momentum-dependent choice;
  \item \code{constant}: a fixed average separation energy, $30\MeV$ by default;
  \item \code{shell}: the argon-40 single-particle levels, each Gaussian-smeared
        about its separation energy and weighted by occupancy.
\end{itemize}
Nucleons drawn from a short-range-correlated pair are not described by a mean
field and always take the two-body recoil energy of Eq.~\eqref{eq:esrc} instead,
whichever prescription is selected; the spectral-function models of
Sec.~\ref{sec:spectral} supply their own joint momentum and removal-energy
distribution and ignore the flag altogether. We describe the default in detail
below, since it is the one used throughout Sec.~\ref{sec:results}, and quantify
the difference between the three prescriptions in Sec.~\ref{sec:results}.

For the mean-field nucleons the default takes the real part of the
momentum-dependent optical potential of Ref.~\cite{Juszczak2005},
\begin{equation}
  V(\kf,p) = -\,\frac{(a\,\kf)^2\,(\kf + b)}
                     {c^4 + d^3\,\kf + e^3\,p^2/\kf + p^4},
  \label{eq:potential}
\end{equation}
with $a=206\MeV$, $b=582\MeV$, $c=-322\MeV$, $d=422\MeV$, and $e=289\MeV$. 
It is attractive, deepest for slow nucleons, and tends
smoothly to zero at high momentum; at the argon Fermi momentum it reaches
$V(p\!=\!0) = -59\MeV$, consistent with Ref.~\cite{Juszczak2005}.

The potential fixes the energy of the bound nucleon $N$ in the nuclear rest
frame, $E_N = \sqrt{p^2 + M_N^2} + V(\kf,p)$, which in turn defines an effective
removal energy $E_{\mathrm{rem}} = M_N - E_N$ and an off-shell invariant mass
$W^2 = E_N^2 - p^2$ that is the true energy budget of the decay. Here $p$ is
always the nucleon momentum and $N$ the bound nucleon, proton or neutron. For the
correlated nucleons the removal energy is set instead by the recoil of the
correlated partner,
\begin{equation}
  E_{\mathrm{rem}} \simeq E_{0} + \frac{p^2}{2M_N},
  \label{eq:esrc}
\end{equation}
where $E_{0}=20\MeV$ is a fixed offset; the same $E_{0}$ sets the correlated
ridge of the effective spectral function in Sec.~\ref{sec:spectral}. When the
binding and Fermi motion drive $W$ below the summed daughter mass, the
decay is kinematically impossible and the nucleon is resampled.

\begin{figure*}[tbp]
  \centering
  \includegraphics[width=0.92\textwidth]{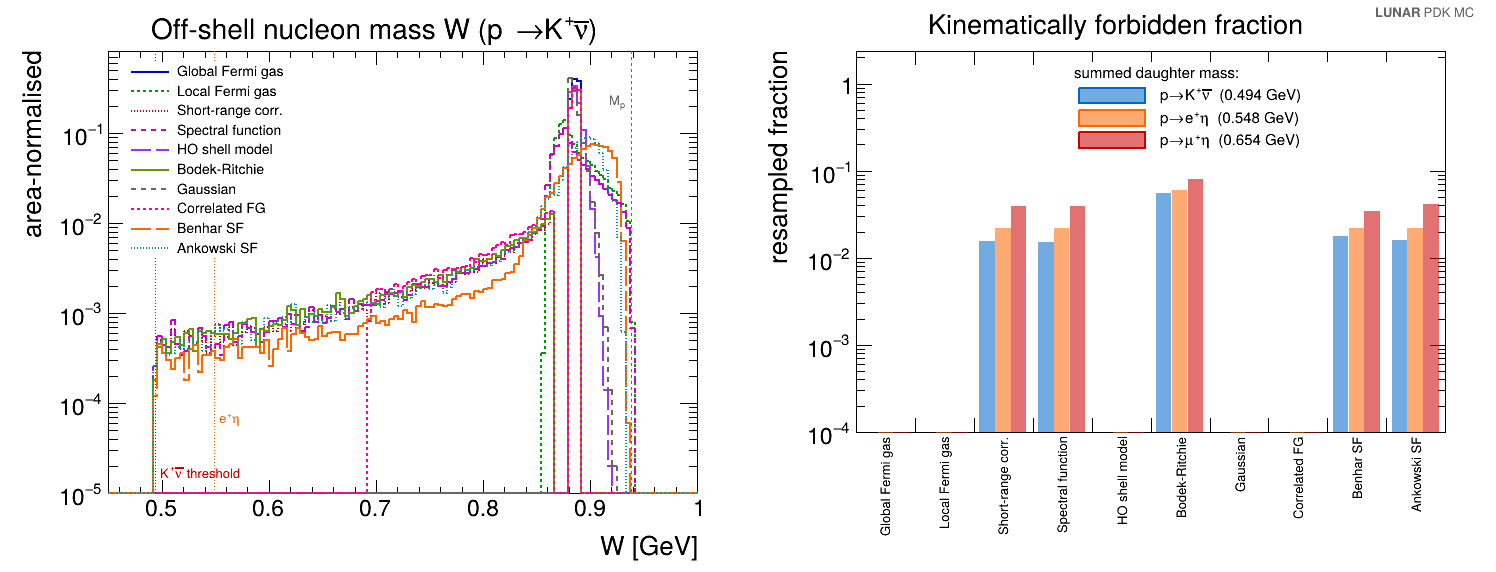}
  \caption{Left: the off-shell nucleon invariant mass
  $W=\sqrt{(M_N-E_{\mathrm{rem}})^2-p^2}$ for $p\to K^{+}\bar\nu$, by momentum
  model (area-normalized, logarithmic scale); the dashed line marks the free
  proton mass and the dotted lines the $K^{+}\bar\nu$ and $e^{+}\eta$ thresholds.
  Right: the resulting
  kinematically forbidden (resampled) fraction per model, for three channels of
  increasing summed daughter mass (logarithmic scale; the mean-field models
  forbid nothing at all and so have no visible bar). Binding pulls $W$ well
  below $M_p$ and the correlated tails feed the low-$W$ region; the loss is
  confined to the models that populate it, and grows with the threshold.}
  \label{fig:offshellW}
\end{figure*}

Figure~\ref{fig:offshellW} shows the off-shell mass distribution for the kaon
mode across the nuclear models, together with the forbidden fraction it implies.
Binding and Fermi motion shift the whole distribution below the free mass $M_p$,
and the correlated and spectral-function models develop a low-$W$ tail fed by their
deeply bound, high-momentum nucleons. That tail is the whole story for the
forbidden fraction, and the right-hand panel quantifies it for three channels of
increasing summed daughter mass. The mean-field models (global and local Fermi
gas, harmonic-oscillator shell model, Gaussian, and the correlated Fermi gas,
whose tail is cut at $2\kf$) never reach below the threshold of any of the three
channels and lose nothing at all. The models with an extended $1/p^4$ tail do: for
$p\to K^{+}\bar\nu$ (threshold $0.494\GeV$) they forbid $1.5$--$1.8\%$ of draws,
rising to $5.5\%$ for Bodek--Ritchie, whose tail extends to $1\GeV/c$. Raising the
threshold raises the loss monotonically: $2.2\%$ ($6.1\%$ for Bodek--Ritchie) for
$p\to e^{+}\eta$ at $0.548\GeV$, and $3.5$--$4.1\%$ ($8.0\%$) for
$p\to\mu^{+}\eta$ at $0.654\GeV$, the heaviest two-body final state in the channel
list. The effect is therefore small in absolute terms but entirely
model-dependent, and it is one more reason to quote nuclear-model spreads rather
than a single prediction. Because forbidden draws are resampled, the spectra
shown throughout this paper are conditioned on the decay being kinematically
allowed; read as an absolute in-medium rate, a channel would be suppressed by the
corresponding few percent. We do not fold this into the yields of
Sec.~\ref{sec:predictions}, where it is well inside the nuclear-model band.

\subsection{Spectral functions}
\label{sec:spectral}

The \code{benhar} and \code{ankowski} options bypass the analytic momentum
sampling entirely and instead draw the momentum and removal energy \emph{jointly}
from a spectral function $S(p,E)$: the momentum from the $p^2$-weighted marginal
$\int S(p,E)\,dE$, and the removal energy from $S(p,\cdot)$ conditioned on it.
Because $S(p,E)$ already encodes both the mean-field shell structure and the
correlated tail in the full $(p,E)$ plane, no separate high-momentum tail or
binding model is layered on top. The two options differ in how $S(p,E)$ is
obtained.

\begin{figure*}[tbp]
  \centering
  \includegraphics[width=0.92\textwidth]{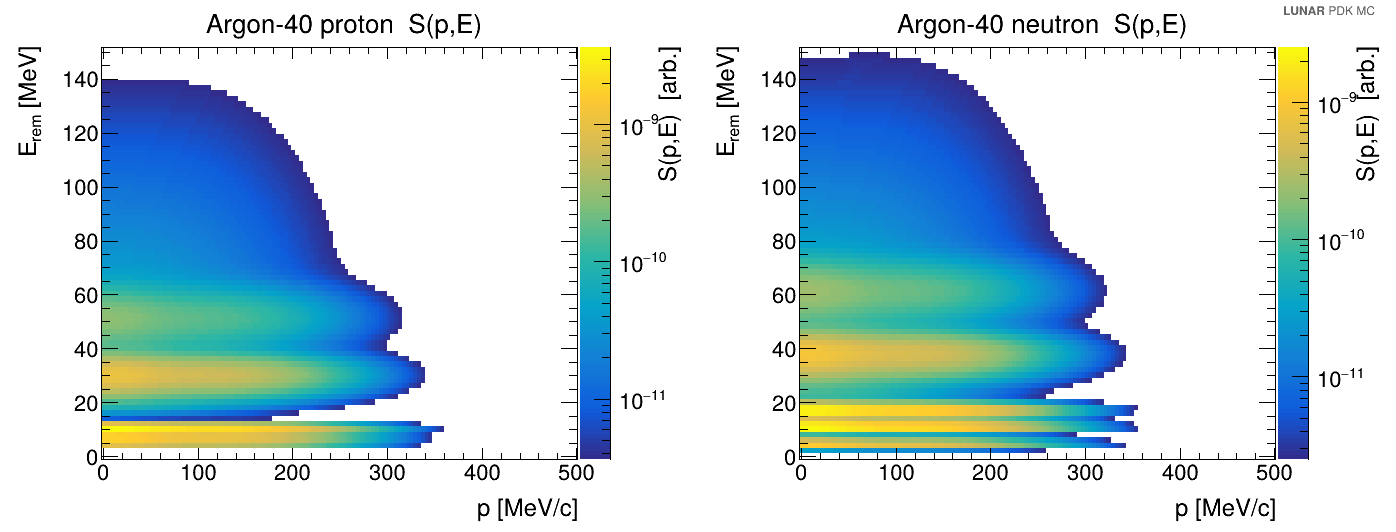}
  \caption{Tabulated argon-40 spectral function $S(p,E)$ (logarithmic color
  scale) for the proton (left) and neutron (right) grids~%
  \cite{JLabArgonSF2022,BanerjeeAnkowski2024} that drive the
  \code{benhar} model. The bright low-$(p,E)$ region is the
  mean-field shell structure; the diffuse high-momentum, high-removal-energy
  strength is the short-range-correlated tail.}
  \label{fig:spectral}
\end{figure*}

The \code{benhar} option reads a tabulated grid. The default grids are the
Jefferson Lab E12-14-012 argon proton and neutron spectral
functions~\cite{JLabArgonSF2022,BanerjeeAnkowski2024} as distributed with
NuWro~\cite{NuWro2025}, chosen automatically according to the parent nucleon.
They are not hard-wired: the \code{--sf-file} option replaces the built-in choice
with \emph{any} tabulated $S(p,E)$ supplied by the user, so a different argon
calculation, or a grid for another nucleus, can be used without touching the
code. The expected format is the plain-text NuWro \code{gridfun2d} layout: a
three-line header giving the number of removal-energy and momentum points and
the two axis ranges in MeV, followed by one block per momentum value listing the
$(E, S)$ pairs (Appendix~\ref{app:usage}). Only relative values matter, and the
sampler builds its marginal and conditional distributions from whatever grid it
is given.
Figure~\ref{fig:spectral} displays the default grids
themselves: a bright mean-field band at low momentum and low removal energy, and a
diffuse correlated ridge reaching to high $p$ and high $E$.

\begin{figure*}[tbp]
  \centering
  \includegraphics[width=0.92\textwidth]{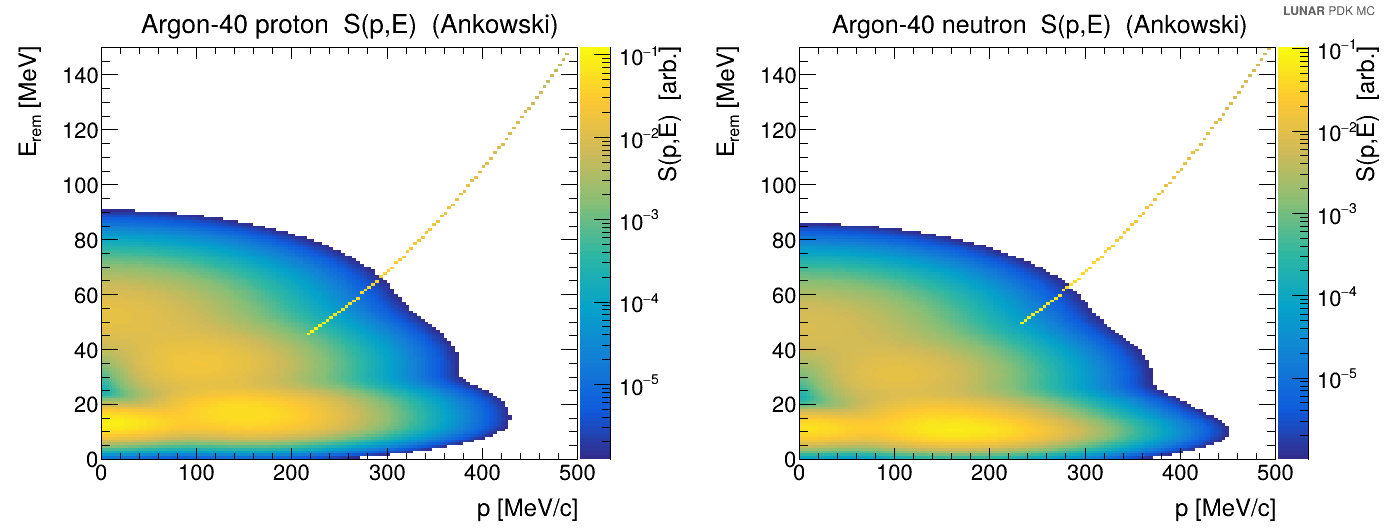}
  \caption{Analytic Ankowski--Sobczyk effective spectral function $S(p,E)$
  (logarithmic color scale) built by the \code{ankowski} model for the proton
  (left) and neutron (right), with no grid input. The horizontal band is the
  shell mean field at the orbital separation energies; the thin ridge climbing
  with momentum is the correlated $1/p^4$ tail at
  $E_{\mathrm{rem}}=E_0+p^2/2M_N$, Eq.~\eqref{eq:esrc}. The neutron
  band reaches to higher momentum through its occupied $1f_{7/2}$ orbital.
  Compare with the tabulated grid of Fig.~\ref{fig:spectral}.}
  \label{fig:ankowskisf}
\end{figure*}

The \code{ankowski} option instead constructs an \emph{effective}
Ankowski--Sobczyk spectral function~\cite{AnkowskiSobczyk2008} analytically, with
no external grid. A
fraction $f_{\mathrm{corr}}=0.20$ of nucleons populate a correlated $1/p^4$ tail
carrying the two-nucleon removal energy $E_{\mathrm{rem}}=E_0+p^2/2M_N$; the remaining nucleons
form a shell mean field, in which each argon orbital contributes its
harmonic-oscillator momentum profile $|\phi_{nl}(p)|^2$ paired with a
Gaussian-broadened separation energy from the $(e,e'p)$ shell table
(Table~\ref{tab:models}'s \code{hosm} profiles and the \code{shell} separation
energies, combined into one joint distribution). The two parts are normalized so
that the correlated strength is exactly $f_{\mathrm{corr}}$ in
$\int 4\pi p^2 S\,dp\,dE$. The resulting $S(p,E)$, shown in
Fig.~\ref{fig:ankowskisf}, reproduces the same qualitative anatomy as the
tabulated grid (a shell band stacked at the orbital separation energies plus a
correlated ridge climbing with momentum) while remaining fully analytic and
grid-free, and gives a measurably broader nucleon-momentum distribution than the
\code{benhar} grid (Table~\ref{tab:momsummary}).

Two features of this construction deserve comment. The value
$f_{\mathrm{corr}}=0.20$ is not tuned here: it is the standard correlated
strength of the effective spectral function of
Ref.~\cite{AnkowskiSobczyk2008}, and it agrees with the $\sim$20\% of nucleons
that many-body calculations place outside the mean field in medium and heavy
nuclei~\cite{BenharFabrociniFantoni1994} and with the short-range-correlated
fraction extracted from electron scattering~\cite{Subedi2008,Fomin2012}. The
sharp ridge that this produces in the $(p,E)$ plane is not a physical feature of
the nucleus but an artifact of the
approximation: the correlated pair is treated as a nucleon recoiling against a
single spectator at rest, so all of its removal-energy strength collapses onto
the line $E_{\mathrm{rem}}=E_0+p^2/2M_N$. In a full calculation the centre-of-mass motion of the
pair and the excitation spectrum of the residual $A-2$ system smear that line
into the broad ridge visible in the tabulated grid of Fig.~\ref{fig:spectral}.
The two options are provided precisely so that this can be tested: the effective
form is fast, analytic and transparent, the tabulated grid carries the realistic
width, and comparing them bounds what the idealization costs. As
Sec.~\ref{sec:results} shows, the resulting kaon spectra differ by only a few
$\MeV/c$ in mean and RMS, because the observable integrates over the ridge.

\subsection{Off-shell two-body decay and channels}
\label{sec:channels}

Given the off-shell mass $W$, a generic two-body decay $N\to d_1 + d_2$ is
generated relativistically. With the K\"all\'en function
$\lambda(a,b,c)=a^2+b^2+c^2-2(ab+bc+ca)$, the daughter momentum in the nucleon
rest frame is
\begin{equation}
  p^{*} = \frac{\sqrt{\lambda(W^2,m_1^2,m_2^2)}}{2W},
  \label{eq:kallen}
\end{equation}
the two daughters are emitted back-to-back with an isotropic orientation, and
both four-vectors are boosted to the laboratory frame with the nucleon velocity
$\boldsymbol\beta=\mathbf{p}/E$. For the default $p\to K^{+}\bar\nu$ mode the
massless antineutrino reduces Eq.~\eqref{eq:kallen} to
$p^{*}=(W^2-m_K^2)/2W$. The nucleon masses are $M_p=0.93827\GeV$ and
$M_n=0.93957\GeV$.

The fourteen supported channels are listed in Table~\ref{tab:channels}. They
cover the standard supersymmetry- and GUT-motivated two-body modes for both bound
protons and neutrons, and almost all of them carry a dedicated Super-Kamiokande
search: the strange modes~\cite{SuperK2005susy,SuperK2014,SuperK2022mK0}, the
$\pi^0$ modes~\cite{SuperK2020epi0}, the $\bar\nu\pi$
modes~\cite{SuperK2014nupi}, the charged-antilepton-plus-meson
modes~\cite{SuperK2017lmeson}, and the $\eta$ modes~\cite{SuperK2024eta}. The
neutron channels reuse the same machinery with the neutron mass, Fermi momentum,
isospin fraction, and shell table. Each mode produces a lepton-side daughter
(an $e^+$, a $\mu^+$, or an invisible $\bar\nu$) and a hadron-side meson whose
laboratory momentum is the primary observable.

\begin{table}[tbp]
  \centering
  \begin{tabular}{@{}llll@{}}
    \toprule
    Key & Mode & Key & Mode \\
    \midrule
    \code{pToKnu}   & $p\to K^{+}\bar\nu$  & \code{nToEPim}  & $n\to e^{+}\pi^{-}$ \\
    \code{pToEPi0}  & $p\to e^{+}\pi^{0}$  & \code{nToMuPim} & $n\to \mu^{+}\pi^{-}$ \\
    \code{pToMuPi0} & $p\to \mu^{+}\pi^{0}$& \code{nToNuPi0} & $n\to \bar\nu\pi^{0}$ \\
    \code{pToNuPip} & $p\to \bar\nu\pi^{+}$& \code{nToNuEta} & $n\to \bar\nu\eta$ \\
    \code{pToEEta}  & $p\to e^{+}\eta$     & \code{nToNuK0}  & $n\to \bar\nu K^{0}$ \\
    \code{pToMuEta} & $p\to \mu^{+}\eta$   & \code{nToEKm}   & $n\to e^{+}K^{-}$ \\
    \code{pToEK0}   & $p\to e^{+}K^{0}$    & & \\
    \code{pToMuK0}  & $p\to \mu^{+}K^{0}$  & & \\
    \bottomrule
  \end{tabular}
  \caption{Nucleon-decay channels supported by LUNAR, selected with the
  \code{--channel} flag.}
  \label{tab:channels}
\end{table}

\subsection{Final-state interactions}
\label{sec:fsi}

The decay meson is born inside the nucleus and must work its way out, and on the
way it interacts with the spectator nucleons. To capture this, LUNAR transports
the decay hadron and every secondary it creates through a semi-classical
intranuclear cascade, switched on by default. The cascade follows the approach of
NuWro's \code{kaskada}~\cite{NuWro2025,GolanFSI2012}, itself descended from the
classic intranuclear-cascade calculations~\cite{Metropolis1958a,Metropolis1958b}
and the pion-transport simulations of Salcedo, Oset and
collaborators~\cite{SalcedoOset1988,OsetSalcedo1987}, and shares its philosophy
with the cascade final-state models of modern neutrino
generators~\cite{AndreopoulosGENIE2010}.

The decay happens where the decaying nucleon is. For the local-density models,
which sample a radius $r$ from the $r^2\rho(r)$ weight of Eq.~\eqref{eq:density}
in order to evaluate $\kf(r)$, the cascade reuses \emph{that} radius as the decay
vertex, so that the momentum, the binding and the vertex all refer to the same
point of the nucleus; only the direction of the vertex is drawn afresh. The
remaining models sample the nucleon momentum from a global distribution and
therefore define no radius; for them the vertex is drawn independently from the
same $r^2\rho(r)$ weight, which reproduces the correct marginal vertex
distribution but not its correlation with the nucleon momentum. The distinction
matters little in practice (the escaping-meson yields of
Sec.~\ref{sec:fsiresults} move by less than a percentage point when the
correlation is switched off) because the two treatments differ only in that
correlation and not in the distribution of vertices itself.

The hadron is then
advanced in short straight steps of $0.05\,\mathrm{fm}$; at each step the local
mean free path is $\lambda=1/(\rho\,\sigma)$, with $\sigma$ the
hadron--nucleon total cross section resolved into its proton and neutron parts.
When an interaction occurs a target nucleon is drawn from a local Fermi sphere of
radius $\kf^{q}(r)$, Eq.~\eqref{eq:localkf}, evaluated at the current position
and for the species $q$ of the struck nucleon. The spectators are thus always
treated as a local Fermi gas, independently of the model chosen for the
\emph{decaying} nucleon: the momentum models of Table~\ref{tab:models} describe
the single nucleon that decays, including its short-range-correlated tail, while
the medium the daughter meson traverses is the mean field. A
sub-process (elastic scattering, charge exchange, absorption, or multi-meson
production) is chosen in proportion to its partial cross section, the final state
is generated in the projectile--target rest frame, and Pauli blocking is imposed
on every outgoing nucleon. Secondaries are cascaded until all particles have
either escaped or been absorbed. Four-momentum and electric charge are conserved
exactly at every vertex, and the leptons, which do not interact strongly, pass
through untouched.

Quasi-elastic and charge-exchange scattering are forward-peaked,
$\mathrm{d}\sigma/\mathrm{d}t\propto e^{Bt}$, with diffractive slopes
$B=4$, $3$ and $5\ \mathrm{GeV^{-2}}$ for $\pi N$, $KN$ and $NN$ respectively.
These slopes are \emph{representative} rather than fitted: they are single
constants of the right order for each system, rather than functions of energy
fitted channel by channel to angular-distribution data. Their provenance is the
standard compilations of the forward-elastic slope parameter $B$ obtained from
fits of $\ln(\mathrm{d}\sigma/\mathrm{d}t)$ to $A+Bt$: Ref.~\cite{Lasinski1972}
for $\pi^{\pm}p$, $K^{\pm}p$, $pp$ and $\bar{p}p$, and Ref.~\cite{Lasinski1969}
for the momentum dependence of $B$ in $\pi^{\pm}p$ and $K^{\pm}p$ down to
$p_{\text{lab}}\simeq0.3\GeV/c$. The constants adopted here reproduce the
ordering $B_{KN}<B_{\pi N}<B_{NN}$ found in those fits, and are deliberately
taken at the low end of the compiled range: the slopes fall well below their
asymptotic high-energy values at the sub-GeV momenta that dominate the cascade,
where the forward peak is weak. At the momenta relevant here the peaking factor
$\exp[2Bp^{*2}(\cos\theta^{*}-1)]$ is close to unity, so the scattering is in any
case nearly isotropic and the choice has little leverage; it redistributes the
direction of an already-counted interaction and leaves the integrated survival
probability unchanged. Absorption and production are treated as isotropic with an
approximate sequential phase space. The model is
designed for spectrum and efficiency studies rather than as a precision transport
code.

A freshly produced meson is not yet a fully formed color singlet and cannot
re-interact at the full hadronic cross section the moment it is created. The
cascade therefore offers an optional formation zone: each produced meson
free-streams over a formation length $L_f=(p/m)\,c\tau_f$ before its interactions
are enabled, with $p/m=\beta\gamma$ dilating the rest-frame formation time
$c\tau_f$ to the laboratory frame, the prescription used in NuWro's
cascade~\cite{NuWro2025,GolanFSI2012}. The formation zone is off by default; when
enabled, the default $c\tau_f=0.342\,$fm follows the SKAT parametrization, and it
suppresses the first interaction of the primary meson and of mesons made in
inelastic production, leaving struck recoil nucleons untouched. Because it can
only recover yield that the cascade was removing, its effect is strongly
channel dependent; we scan it in Sec.~\ref{sec:fsiresults} and use that scan as a
proxy for the spread between FSI models.

\begin{figure*}[tbp]
  \centering
  \includegraphics[width=0.92\textwidth]{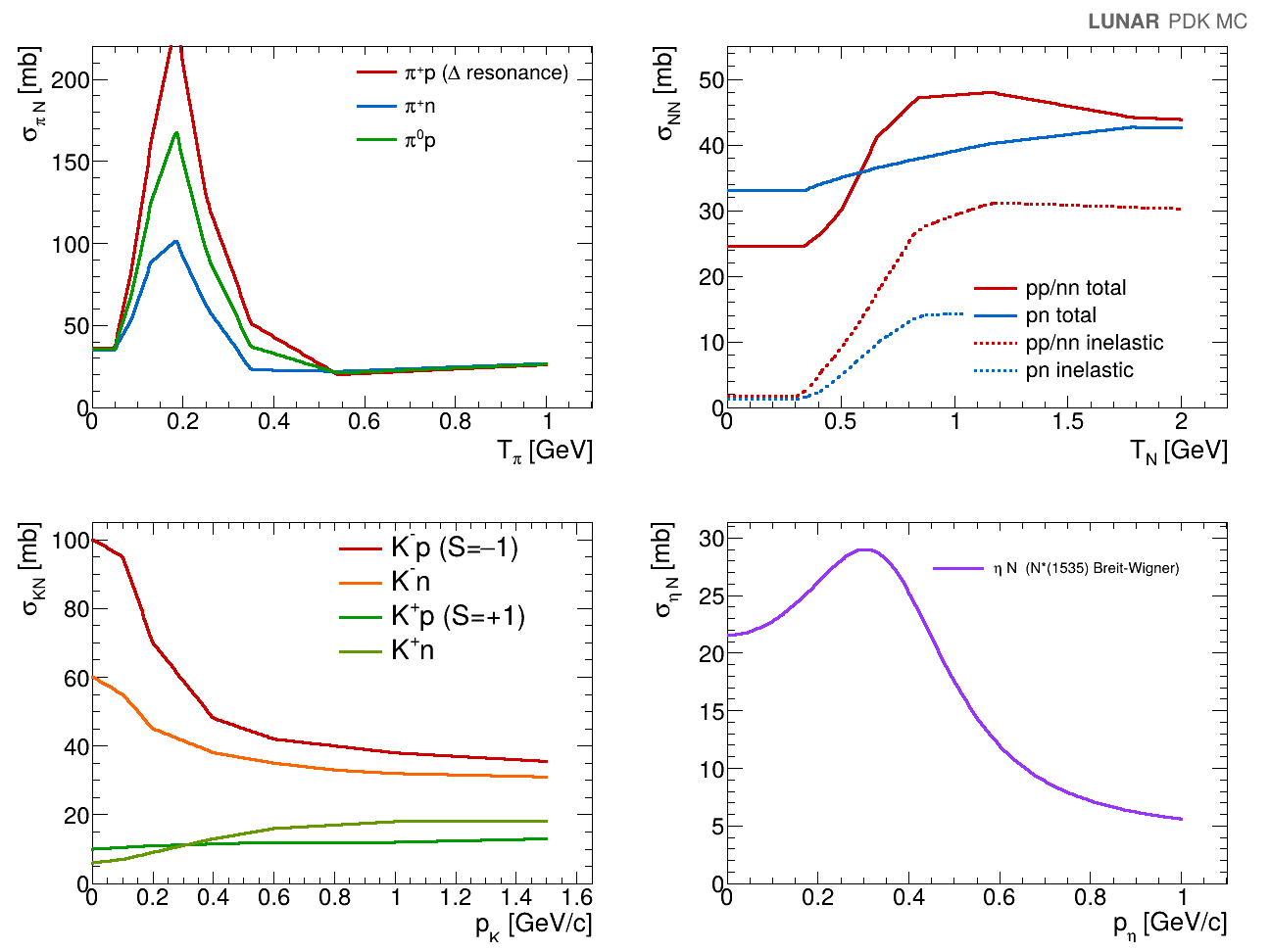}
  \caption{Hadron--nucleon cross sections used by the LUNAR cascade: the
  $\Delta(1232)$ peak in $\pi N$ (top left), the onset of $NN$ inelasticity
  (top right), the nearly transparent $K^{+}$ contrasted with the strongly
  absorbed $K^{-}$ (bottom left), and the $N^{*}(1535)$ resonance in $\eta N$
  (bottom right).}
  \label{fig:fsixsec}
\end{figure*}

The cross sections the cascade uses are shown in Fig.~\ref{fig:fsixsec}. For
pions and nucleons we adopt the Metropolis tables~%
\cite{Metropolis1958a,Metropolis1958b} as used in NuWro. The strange and $\eta$
systems have no such tabulation, so we build short interpolation tables from the
measured totals in the same spirit. The $S=+1$ kaons ($K^{+}$, $K^{0}$) are
nearly transparent in nuclear matter, with total cross sections of only
$\sim$10--18~mb and charge exchange their sole inelasticity; the $S=-1$
antikaons interact strongly ($\sim$30--100~mb, rising toward threshold) and are
dominated by absorption through $\bar K N\to Y\pi$~%
\cite{DoverWalker1982,FriedmanGal2007}. The $\eta N$ cross section is modeled by
the $N^{*}(1535)$ Breit--Wigner that sits directly above the production
threshold, with $\eta N\to\pi N$ as the leading inelastic channel.

A neutral kaon is propagated as the strangeness eigenstate in which it is
produced, not as a $K^0_S/K^0_L$ or an incoherent $K^0$--$\bar K^0$ mixture. The
decays $p\to\ell^{+}K^{0}$ and $n\to\bar\nu K^{0}$ emit a $|K^{0}\rangle$, which
carries $S=+1$ and therefore interacts with the same nearly transparent $S=+1$
cross sections as the $K^{+}$; a $\bar K^{0}$ appears only if charge exchange
makes one. This is the physically correct choice inside the nucleus: the strong
interaction conserves strangeness, and the kaon crosses the residual nucleus in
$\sim$$10^{-23}\,$s, some twelve orders of magnitude faster than the $K^0_S$
lifetime, so the weak $K^0$--$\bar K^0$ mixing that eventually turns the produced
state into a $K^0_S/K^0_L$ superposition has no time to develop while the meson
is still inside the medium. Mixing and decay are applied afterwards, outside the
nucleus, when the escaping neutral kaon is optionally decayed. The consequence is
quantitative and not small: with the incoherent-mixture approximation half of the
neutral-kaon flux is exposed to $\bar KN\to Y\pi$ absorption and the predicted
$K^{0}$ survival drops to $\sim$60\%, whereas the eigenstate treatment leaves the
$K^{0}$ as transparent as the $K^{+}$ (Sec.~\ref{sec:fsiresults}).

LUNAR ships a single cascade, and it is worth being explicit about what that
means for the uncertainty budget, since final-state interactions turn out to be
the leading correction for most channels. Other approaches to the same physics
differ substantially in construction: the intranuclear cascades of GENIE
(\code{hA} and \code{hN})~\cite{AndreopoulosGENIE2010} and NEUT trade a
data-driven effective treatment against a microscopic one, the Li\`ege cascade
INCL++~\cite{Mancusi2014} propagates all participants simultaneously with a
mean-field potential, and GiBUU~\cite{BussGiBUU2012} replaces the cascade
altogether by a Boltzmann--Uehling--Uhlenbeck transport with in-medium spectral
functions and potentials, the approach used in the proton-decay study of
Ref.~\cite{Yan2026}. A meaningful comparison would require running the same
decays through those codes, which is beyond the scope of a dedicated generator
paper. We instead bound the sensitivity from within LUNAR in two ways: by
benchmarking the absorption probabilities against independent treatments
(Table~\ref{tab:fsibench}), and by scanning the formation zone
(Table~\ref{tab:fz}), which is the single ingredient that moves the survival
probabilities most and which we carry as an explicit FSI-model band on the
predicted event rates in Sec.~\ref{sec:predictions}. Interfacing LUNAR to an
external cascade through a standard event record (Sec.~\ref{sec:extensions})
would make the direct comparison straightforward, and we regard it as the natural
next step.

\section{Results}
\label{sec:results}

\subsection{Daughter spectrum: Fermi motion versus binding}

The laboratory momentum of the decay meson is the experimental signature. A free
proton at rest emits a kaon of fixed momentum
$p_K^{\mathrm{free}}=(M_p^2-m_K^2)/2M_p = 0.339\GeV/c$; in the nucleus, Fermi
motion broadens this line and binding shifts it down through the reduced off-shell
mass. The generator lets us vary one at a time and see which attribute of the
spectrum each controls. All numbers in this subsection come from samples of
$2\times10^5$ decays per configuration, so the statistical uncertainty on a mean
is well below $1\MeV/c$ and every difference quoted is a model difference.

\begin{figure}[tbp]
  \centering
  \includegraphics[width=\columnwidth]{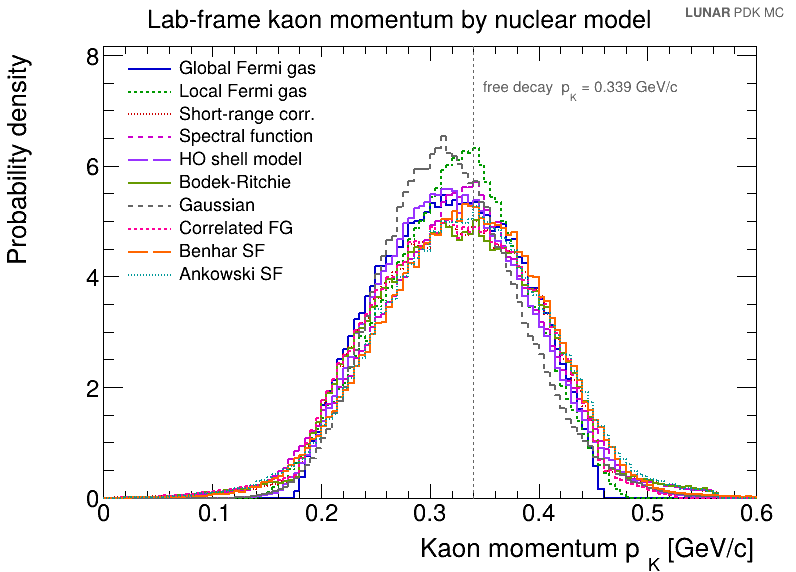}
  \caption{Laboratory kaon-momentum spectrum for the ten physics momentum models
  (area-normalized), all with the default optical-potential binding; the dashed
  line marks the free-decay value
  $p_K=0.339\GeV/c$. The pure Fermi-gas models give the narrowest peaks, while
  the correlated and spectral-function models broaden the spectrum and feed a
  low-momentum shoulder. The analytic \code{ankowski} spectral function is
  slightly broader than the tabulated \code{benhar} grid. Line styles as well as
  colors distinguish the models; models sharing a style differ in color.}
  \label{fig:kaonmodels}
\end{figure}

Varying the momentum model (Fig.~\ref{fig:kaonmodels}) changes the spectral
width. The mean-field models (the global and local Fermi gases, the
harmonic-oscillator shell model, and the Gaussian) give the narrowest kaon
peaks, with a root-mean-square spread of $63$--$67\MeV/c$. The correlated and
spectral-function models broaden it to $73$--$77\MeV/c$, as their high-momentum tails
feed a soft shoulder that pushes as much as $\sim$18\% of kaons below
$0.25\GeV/c$. The Bodek--Ritchie and correlated-Fermi-gas options, which carry
the most strength above $\kf$ in Table~\ref{tab:momsummary}, produce the largest
low-momentum tails, a direct mapping of initial-state high-momentum content onto
the observable. The momentum model also moves the central value: the mean kaon
momentum runs from $0.320\GeV/c$ for the Gaussian to $0.330\GeV/c$ for the two
spectral functions, a spread of about $10\MeV/c$.

\begin{figure*}[tbp]
  \centering
  \includegraphics[width=0.92\textwidth]{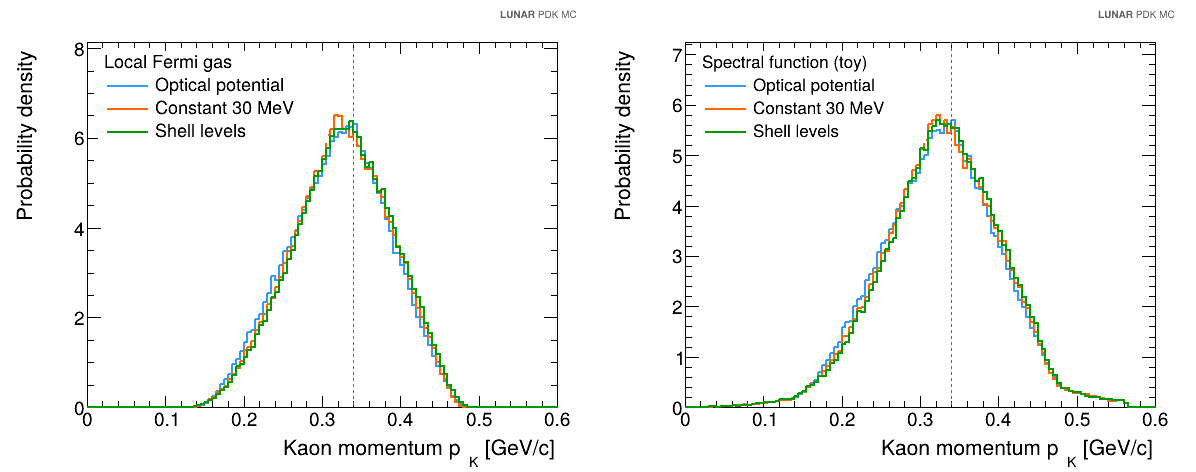}
  \caption{Kaon-momentum spectrum across the three binding models, for the local
  Fermi gas (left) and the toy spectral function (right); the dashed line marks
  the free-decay value. The binding choice shifts the peak almost rigidly, the
  optical potential, being the deepest, giving the softest kaon and the broadest
  low-momentum shoulder.}
  \label{fig:kaonbinding}
\end{figure*}

Varying the binding model instead (Fig.~\ref{fig:kaonbinding} and
Table~\ref{tab:kaonbinding}) shifts the spectrum rigidly with almost no change in
shape. The optical potential, being the deepest, yields the softest kaon; the
constant and shell prescriptions lie close together and give slightly harder
spectra. The
displacement of the mean is $\sim$6$\MeV/c$, while the root-mean-square is
fixed to better than $1\MeV/c$.

The two nuclear ingredients are therefore \emph{not} interchangeable in their
effects, but neither are they orthogonal. The width of the kaon spectrum is
controlled by the momentum model alone: the binding prescription leaves the RMS
unchanged, while going from a mean-field to a correlated or spectral-function
momentum distribution widens it by $\sim$20\%. The central value responds to
both, and, importantly, it responds more strongly to the
momentum model ($\sim$10$\MeV/c$ across the ten models) than to the binding
prescription ($\sim$6$\MeV/c$ across the three). A meaningful nuclear
uncertainty on the mean must therefore vary the momentum model; varying only the
binding prescription, the more obvious knob, understates it by roughly a factor
of two. What can be said cleanly is the asymmetry: the binding model affects the
central value and nothing else, whereas the momentum model affects everything.

\begin{table}[tbp]
  \centering
  \footnotesize
  \setlength{\tabcolsep}{4pt}
  \begin{tabular}{@{}llccc@{}}
    \toprule
    Momentum model & Binding & $\langle p_K\rangle$ & RMS & in-win. \\
                   &         & [\si{GeV/c}] & [\si{GeV/c}] & [\%] \\
    \midrule
    Local Fermi gas    & potential & 0.321 & 0.063 & $45.4$ \\
                       & constant & 0.325 & 0.062 & $46.0$ \\
                       & shell & 0.327 & 0.063 & $46.0$ \\
    \addlinespace
    Spectral fn.\ (toy)& potential & 0.325 & 0.075 & $41.2$ \\
                       & constant & 0.328 & 0.074 & $41.7$ \\
                       & shell & 0.330 & 0.074 & $41.9$ \\
    \bottomrule
  \end{tabular}
  \caption{Laboratory kaon momentum versus binding model for two analytic
  momentum models ($2\times10^5$ events each). ``In-window'' is the fraction of
  decays whose kaon momentum falls inside
  $[0.30,0.38]\GeV/c$, i.e.\ within $\pm0.04\GeV/c$ of the free-decay value, the
  containment proxy defined in Sec.~\ref{sec:predictions}. The binding choice
  moves the mean by $\sim$6$\MeV/c$ and leaves the RMS fixed, whereas changing
  the momentum model changes the RMS by $\sim$20\%. The spectral-function models
  are binding-independent by construction.}
  \label{tab:kaonbinding}
\end{table}

\subsection{Final-state interactions, channel by channel}
\label{sec:fsiresults}

The cascade reshapes the observable spectrum and, more importantly, removes
signal mesons. We classify each event by the fate of the primary meson: it may
escape untouched (\code{none}), scatter elastically, emerge as a different meson
through charge exchange (\code{cex}), survive while spawning additional mesons
(\code{produced}), or fail to escape (\code{absorbed}). The first, second and
fourth of these leave a meson of the original species in the final state, and
their sum is the quantity that propagates into every event rate we quote:
\begin{equation}
  \varepsilon_{\text{FSI}} \equiv
  P(\code{none}) + P(\code{elastic}) + P(\code{produced}),
  \label{eq:epsfsi}
\end{equation}
the probability that the decay meson escapes the argon nucleus as itself. It is
a survival probability, not a detector efficiency; charge exchange counts against
it because the escaping meson is then a different particle from the one the
search is looking for.

\begin{figure*}[tbp]
  \centering
  \includegraphics[width=0.92\textwidth]{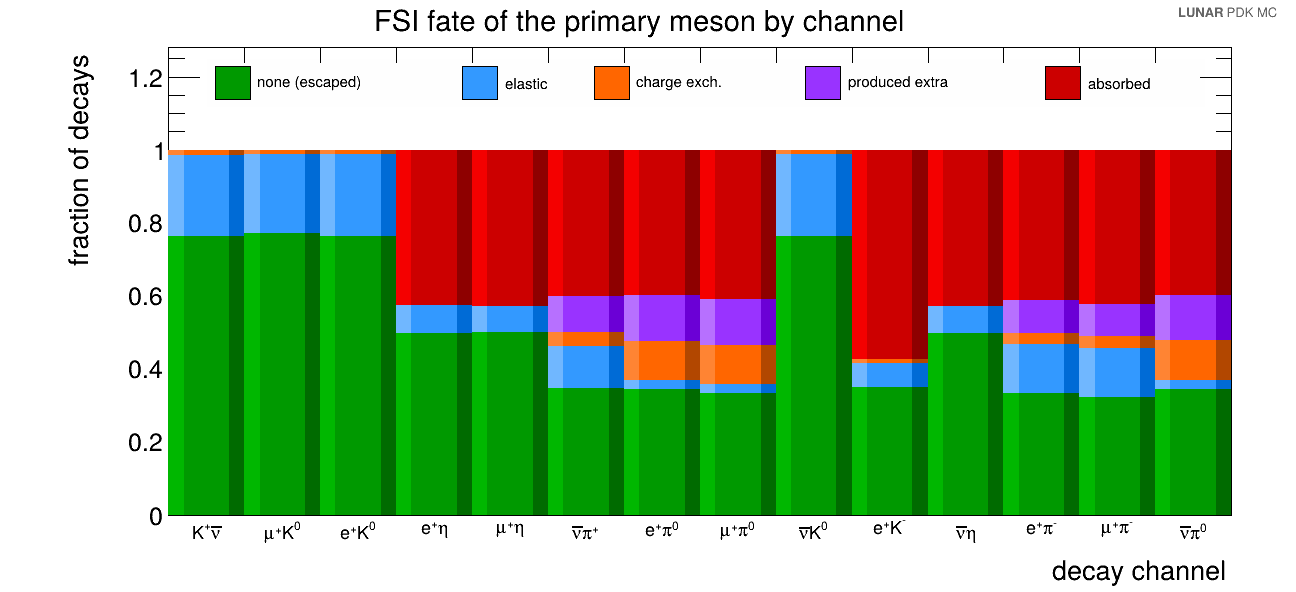}
  \caption{Fate of the primary meson after the cascade in argon, for all fourteen
  proton and neutron channels (each bar normalized to one decay). The $S=+1$
  kaons are nearly transparent; the pions, the $\eta$, and the antikaon are
  strongly absorbed or charge-exchanged. Local Fermi gas, $2\times10^{5}$ events
  for the six benchmark channels of Table~\ref{tab:fsi} and $5\times10^{4}$ for
  the rest.}
  \label{fig:fsioutcomes}
\end{figure*}

Figure~\ref{fig:fsioutcomes} and Table~\ref{tab:fsi} give these fractions for six
benchmark channels, and the contrast between them is the central result of this
section. The $S=+1$ kaons are almost transparent: three quarters escape without
interacting, the rest scatter elastically, and none is absorbed, since a $K^{+}$
or $K^{0}$ has no strangeness-conserving absorption channel on a nucleon. The
pions are heavily reworked, with roughly 40\% absorbed and a further quarter
charge-exchanged or augmented by produced mesons. The $\eta$ and the antikaon are
the most strongly absorbed (43\% and 57\%), the latter because its cross section
grows toward threshold through $\bar K N\to Y\pi$.

\begin{table*}[tbp]
  \centering
  \begin{tabular}{@{}llccccccc@{}}
    \toprule
    Channel & Meson & none & elastic & cex & prod. & abs. & yield & $\langle p\rangle$ shift [\si{GeV/c}] \\
    \midrule
    $p\to K^{+}\bar\nu$  & $K^{+}$    & 0.76 & 0.22 & 0.01 & 0.00 & 0.00 & $99\%$ & $0.321\!\to\!0.295$ \\
    $p\to\mu^{+}K^{0}$   & $K^{0}$    & 0.77 & 0.22 & 0.01 & 0.00 & 0.00 & $99\%$ & $0.307\!\to\!0.282$ \\
    $p\to e^{+}\eta$     & $\eta$     & 0.50 & 0.08 & 0.00 & 0.00 & 0.43 & $57\%$ & $0.293\!\to\!0.279$ \\
    $p\to e^{+}\pi^{0}$  & $\pi^{0}$  & 0.34 & 0.02 & 0.11 & 0.12 & 0.40 & $49\%$ & $0.442\!\to\!0.309$ \\
    $p\to\bar\nu\pi^{+}$ & $\pi^{+}$  & 0.35 & 0.12 & 0.04 & 0.10 & 0.40 & $56\%$ & $0.441\!\to\!0.365$ \\
    $n\to e^{+}K^{-}$    & $K^{-}$    & 0.35 & 0.07 & 0.01 & 0.00 & 0.57 & $42\%$ & $0.321\!\to\!0.311$ \\
    \bottomrule
  \end{tabular}
  \caption{Primary-meson outcome fractions, escaping signal-meson yield, and the
  shift of the mean observable momentum (cascade off~$\to$~on, the latter averaged
  over escaping mesons of the signal species) for six benchmark
  channels in argon ($2\times10^{5}$ events each, local Fermi gas). The yield is
  the survival probability $\varepsilon_{\text{FSI}}$ of Eq.~\eqref{eq:epsfsi},
  used in the event-rate predictions of
  Sec.~\ref{sec:predictions}. Both $S=+1$ kaons are transparent: with strangeness
  conserved by the strong interaction, neither the $K^{+}$ nor the $K^{0}$ has an
  absorption channel on a nucleon.}
  \label{tab:fsi}
\end{table*}

The survival and absorption fractions of Table~\ref{tab:fsi} rest on validated
microphysics rather than free parameters: the $\pi N$ and $NN$ cross sections are
the Metropolis tables~\cite{Metropolis1958a,Metropolis1958b} used by
NuWro~\cite{NuWro2025}, embedded verbatim, and the $K$-- and $\eta$--nucleon
totals reproduce the measured data~\cite{DoverWalker1982,FriedmanGal2007}.
Table~\ref{tab:fsibench} benchmarks the resulting meson absorption in argon
against independent treatments: the $\pi^{0}$ absorption matches the GiBUU
transport result of Yan \textit{et al.}~\cite{Yan2026}, and the $S=+1$ kaons are
unabsorbed, lying below the $\bar K N\to Y\pi$ threshold that makes
the $S=-1$ $K^{-}$ opaque, as in the kaon survey of Barbu \textit{et
al.}~\cite{Barbu2025}. The only ingredients that are not taken from data are the
elastic and charge-exchange angular slopes, which are representative constants
rather than channel-by-channel fits (Sec.~\ref{sec:fsi}); because they only
redistribute the direction of an interaction that has already been counted, the
integrated survival $\varepsilon_{\text{FSI}}$ is insensitive to them.

\begin{table}[tbp]
  \centering
  \begin{tabular}{@{}lccl@{}}
    \toprule
    Meson & $S$ & LUNAR abs. & Independent treatment \\
    \midrule
    $\pi^{0}$ & $0$  & $0.40$ & ${\sim}0.40$ ($^{16}$O, GiBUU)~\cite{Yan2026} \\
    $K^{+}$   & $+1$ & $0.00$ & nearly transparent~\cite{DoverWalker1982,Barbu2025} \\
    $K^{0}$   & $+1$ & $0.00$ & nearly transparent~\cite{DoverWalker1982} \\
    $K^{-}$   & $-1$ & $0.57$ & strongly absorbed~\cite{FriedmanGal2007,Barbu2025} \\
    \bottomrule
  \end{tabular}
  \caption{Benchmark of the primary-meson absorption probability in argon
  (from Table~\ref{tab:fsi}) against independent treatments. Because the
  $\pi N$/$NN$ cross sections are the NuWro Metropolis tables and the $K$/$\eta$
  totals follow measured data, the comparison tests the cascade rather than its
  inputs.}
  \label{tab:fsibench}
\end{table}

Table~\ref{tab:fz} scans the surviving-meson yield against the formation time
$c\tau_f$ of the formation zone described in Sec.~\ref{sec:fsi}. Because the
formation zone can only recover yield that the cascade was
removing, its impact is sharply channel dependent. The $S=+1$ kaon modes are
unchanged ($99\%\to99\%$): the kaon is slow ($\beta\gamma\simeq0.7$,
$L_f\lesssim0.3\,$fm at the default) and already transparent, so there is nothing
to recover. The
strongly-interacting pions shift the most (the $e^+\pi^0$ yield climbs from
$49\%$ with no formation zone to $60\%$ at the default and to $81\%$ at
$1.0\,$fm, and the $\bar\nu\pi^+$ from $56\%$ to $66\%$ and $83\%$) because the
energetic pions ($\beta\gamma\sim3$) free-stream a sizeable fraction of the
nuclear radius before they can be absorbed; the slower $\eta$ and $\bar K$ modes
gain two to three points at the default and up to $\sim$7 percentage
points at $1.0\,$fm. The formation zone is thus a leading systematic for the
pion, $\eta$, and antikaon channels and a negligible one for the kaon
searches. The predictions of Sec.~\ref{sec:predictions} use the default cascade
(formation zone off), which is the conservative end of this range, and we carry
the $0.342$--$1.0\,$fm spread of Table~\ref{tab:fz} as an explicit FSI-model band
on $\varepsilon_{\text{FSI}}$, and hence on the predicted event rates.

\begin{table*}[tbp]
  \centering
  \begin{tabular}{@{}llcccc@{}}
    \toprule
    & & & \multicolumn{3}{c}{$c\tau_{f}$ [\si{fm}]} \\
    \cmidrule(l){4-6}
    Channel & Meson & no FZ & $0.342$ & $0.5$ & $1.0$ \\
    \midrule
    $p\to K^{+}\bar\nu$  & $K^{+}$    & $99\%$ & $99\%$ & $99\%$ & $99\%$ \\
    $p\to\mu^{+}K^{0}$   & $K^{0}$    & $99\%$ & $99\%$ & $99\%$ & $99\%$ \\
    $p\to e^{+}\eta$     & $\eta$     & $57\%$ & $59\%$ & $60\%$ & $63\%$ \\
    $p\to e^{+}\pi^{0}$  & $\pi^{0}$  & $49\%$ & $60\%$ & $66\%$ & $81\%$ \\
    $p\to\bar\nu\pi^{+}$ & $\pi^{+}$  & $56\%$ & $66\%$ & $70\%$ & $83\%$ \\
    $n\to e^{+}K^{-}$    & $K^{-}$    & $42\%$ & $44\%$ & $46\%$ & $49\%$ \\
    \bottomrule
  \end{tabular}
  \caption{Formation-zone sensitivity of the escaping signal-meson yield (the
  survival probability \code{none}$+$\code{elastic}$+$\code{produced}, i.e.\ the
  $\varepsilon_{\text{FSI}}$ used in Sec.~\ref{sec:predictions}) for the six
  benchmark channels, as the formation time $c\tau_f$ is scanned. ``no FZ'' is the
  default cascade (the yield column of Table~\ref{tab:fsi}); the formation zone
  monotonically raises the yield, by $\lesssim$1 point for the transparent $S=+1$
  kaons but by tens of points for the strongly-absorbed pions. The scan is run for
  all fourteen channels and supplies the FSI-model band of
  Sec.~\ref{sec:predictions}; the six benchmarks are shown here. Same sample as
  Table~\ref{tab:fsi} ($2\times10^{5}$ events, local Fermi gas).}
  \label{tab:fz}
\end{table*}

\begin{figure*}[tbp]
  \centering
  \includegraphics[width=0.92\textwidth]{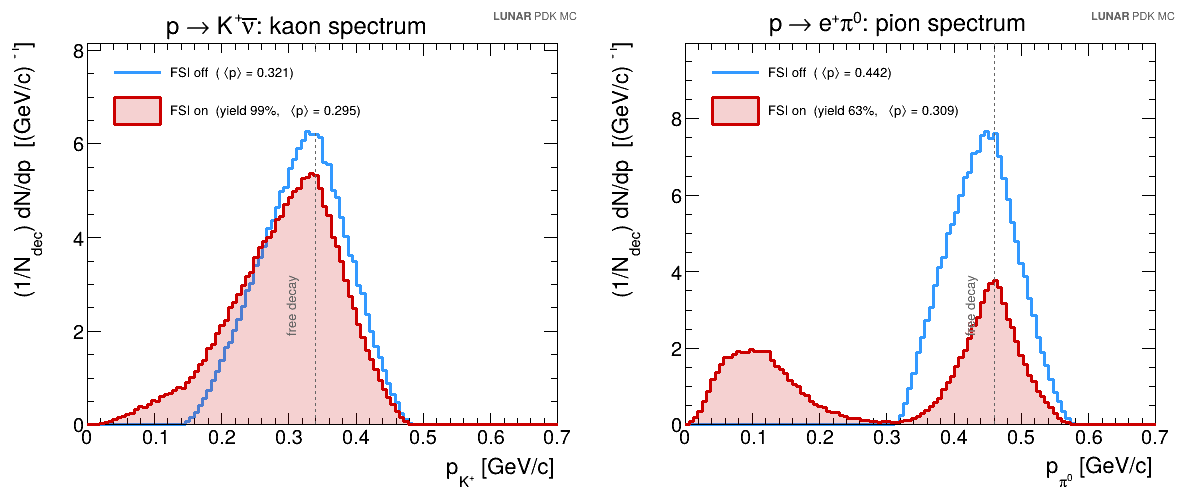}
  \caption{Laboratory meson momentum with (red, filled) and without (blue) the
  cascade, normalized per generated decay, for the transparent $K^{+}$ (left) and
  the strongly interacting $\pi^{0}$ (right); the dashed line marks the free-decay
  value. The $K^{+}$ is barely affected, whereas the $\pi^{0}$ loses about half
  its yield and develops a large low-momentum bump. The legend quotes the integral
  of the ``FSI on'' histogram, which counts every meson of the signal species in
  the final state; for the $\pi^{0}$ this is $0.64$ per decay rather than the
  primary-meson survival probability $\varepsilon_{\text{FSI}}=49\%$ of
  Table~\ref{tab:fsi}, the difference being pions made in the cascade when the
  primary was absorbed or charge-exchanged, which populate the low-momentum bump.
  For the transparent $K^{+}$ the two coincide.}
  \label{fig:fsicompare}
\end{figure*}

\begin{figure*}[tbp]
  \centering
  \includegraphics[width=0.92\textwidth]{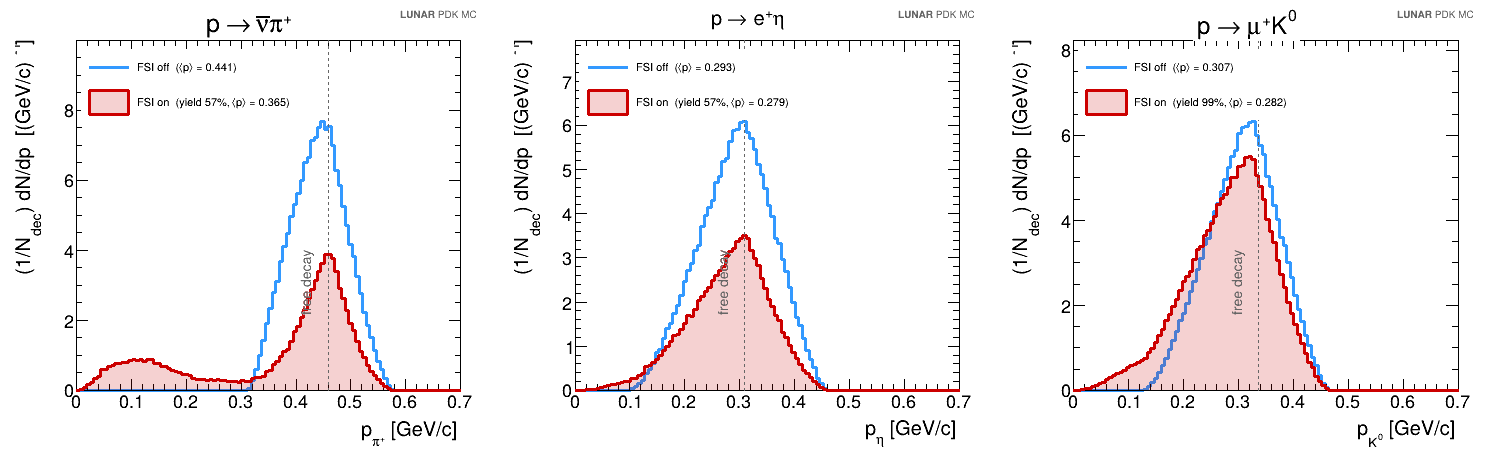}
  \caption{Meson momentum with (red) and without (blue) the cascade for the
  $\pi^{+}$ (left), $\eta$ (middle), and $K^{0}$ (right) channels, normalized per
  generated decay. The $\pi^{+}$ is depleted and softened most strongly and the
  $\eta$ substantially; the $S=+1$ $K^{0}$ keeps essentially all of its yield and
  only softens, tracking the $K^{+}$ of Fig.~\ref{fig:fsicompare}.}
  \label{fig:fsispectra}
\end{figure*}

The effect on the spectrum is shown in Figs.~\ref{fig:fsicompare}
and~\ref{fig:fsispectra}, where every curve is normalized per generated decay so
that the cascade-on histogram exposes both the depletion (a smaller area, from
absorption and charge exchange) and the softening, a low-momentum tail left by
quasi-elastic energy loss. For the signal mode the kaon stays $\sim$99\%
transparent and merely softens slightly, its mean falling from $0.321$ to
$0.295\GeV/c$. The $p\to e^{+}\pi^{0}$ pion, by contrast, drops to a 49\% yield
and grows the pronounced low-momentum bump that is the hallmark of pion
final-state interactions. The same behavior carries over to the $\pi^+$ and
$\eta$ channels, while the $K^0$ follows the $K^{+}$. Across all modes the
surviving-meson yield (the yield column of Table~\ref{tab:fsi}, whose
components are shown channel by channel in Fig.~\ref{fig:fsioutcomes}) ranges
from
$\sim$99\% for the $S=+1$ kaons down to $\sim$42\%
for the strongly absorbed $K^-$: the cascade is a negligible correction for the
kaon searches but a leading effect for every other channel.

\begin{figure*}[tbp]
  \centering
  \includegraphics[width=0.92\textwidth]{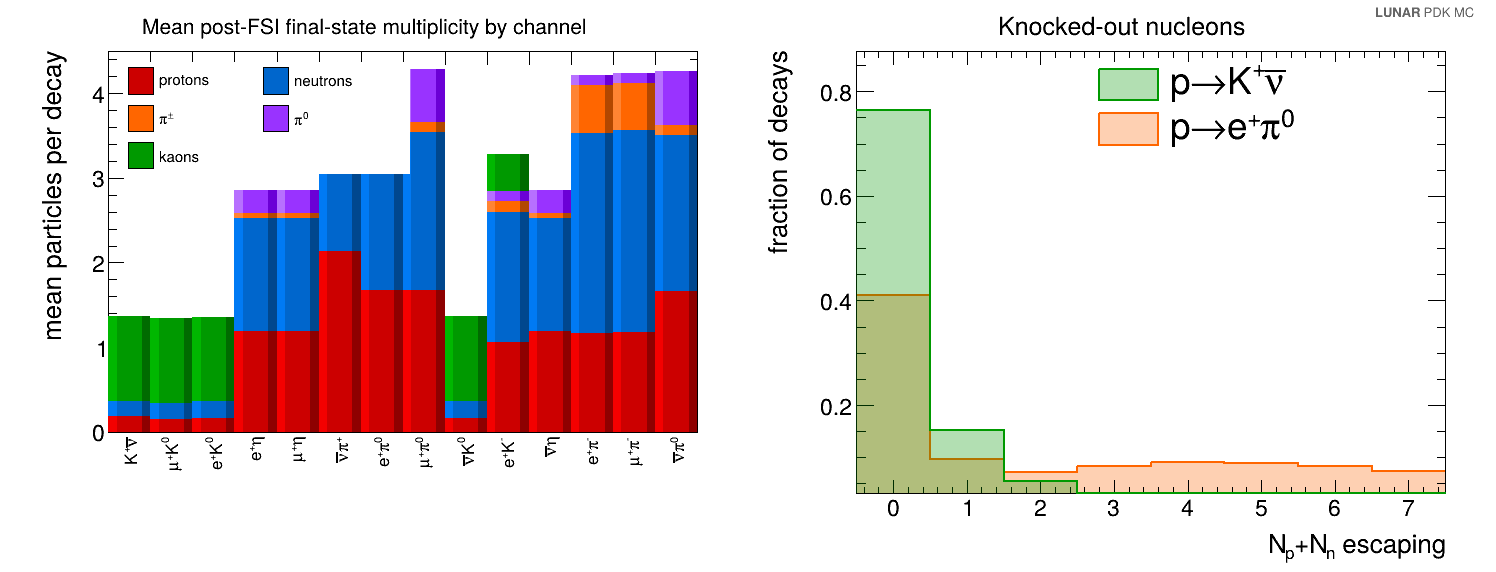}
  \caption{Left: mean number of escaping protons, neutrons, charged and neutral
  pions, and kaons per decay, by channel. Right: the knocked-out nucleon
  multiplicity for the transparent $p\to K^{+}\bar\nu$ and the
  pion-cascade-dominated $p\to e^{+}\pi^{0}$.}
  \label{fig:fsimult}
\end{figure*}

Because the cascade knocks nucleons loose and can add or charge-exchange pions,
the observable final state is richer than a single meson.
Figure~\ref{fig:fsimult} shows the mean post-cascade multiplicity of escaping
protons, neutrons, pions, and kaons per decay. The $K^+$ mode leaves with about
one kaon and almost no extra hadrons, while the pion and $\eta$ modes eject two to
four nucleons and a soft pion halo. This accompanying activity is directly
relevant to event reconstruction in a liquid-argon detector, where additional
short tracks and vertex activity around the decay candidate both aid
identification and complicate it~\cite{Bueno2007}.

\subsection{Expected event yields in DUNE}
\label{sec:predictions}

We now combine the nuclear-model and cascade outputs into expected event counts
for DUNE. For simplicity we assume two 10-kt liquid-argon modules (20~kt fiducial) and the
technical-design exposure of $400\ \mathrm{kt\cdot yr}$, about twenty
years~\cite{DUNE2020}. Argon-40 supplies $2.71\times10^{32}$ protons and
$3.31\times10^{32}$ neutrons per kilotonne, so a proton mode draws on
$N_p=5.4\times10^{33}$ and a neutron mode on $N_n=6.6\times10^{33}$ target
nucleons. The expected count is
\begin{equation}
  N = N_{\text{nuc}}\,\big(1-e^{-T/\tau}\big)\,\varepsilon
    \;\simeq\; N_{\text{nuc}}\,\frac{T}{\tau}\,\varepsilon
    \qquad (\tau\gg T),
  \label{eq:rate}
\end{equation}
with the detection efficiency factorized in the same way for every mode,
\begin{equation}
  \varepsilon = \varepsilon_{\text{det}}\;\varepsilon_{\text{FSI}},
  \label{eq:eps}
\end{equation}
where $\varepsilon_{\text{FSI}}$ is the cascade survival of
Eq.~\eqref{eq:epsfsi}, applied exactly once, and $\varepsilon_{\text{det}}$ is a
flat 30\% reconstruction assumption, the same for all fourteen channels. We
deliberately do not substitute the mode-specific DUNE technical-design
efficiencies quoted for selected channels~\cite{DUNE2020}. Those DUNE numbers are evaluated on a simulation that \emph{already}
includes Fermi motion and final-state interactions, the $p\to e^{+}\pi^{0}$
value being limited by $\pi^{0}$ reabsorption, so they are not
$\varepsilon_{\text{det}}$ in the sense of Eq.~\eqref{eq:eps} and cannot be
multiplied by $\varepsilon_{\text{FSI}}$ without double-counting the cascade. A
single flat
$\varepsilon_{\text{det}}$ keeps the comparison honest, leaves the golden kaon
mode essentially where the DUNE number puts it
($0.30\times0.99\simeq0.30$), and is the more conservative choice for the
$\pi^{0}$ modes. It also means the counts scale trivially: an analyst with a
better per-channel reconstruction efficiency can rescale any row by
$\varepsilon_{\text{det}}/0.30$. For
each mode we set the lifetime $\tau$ to the current Super-Kamiokande 90\%~C.L.\
lower limit on $\tau/B$~\cite{PDG2024}, so that $N$ is the largest yield
consistent with present data, a discovery-reach ceiling that DUNE would fall
below if the true lifetime is longer.

\begin{figure*}[tbp]
  \centering
  \includegraphics[width=0.92\textwidth]{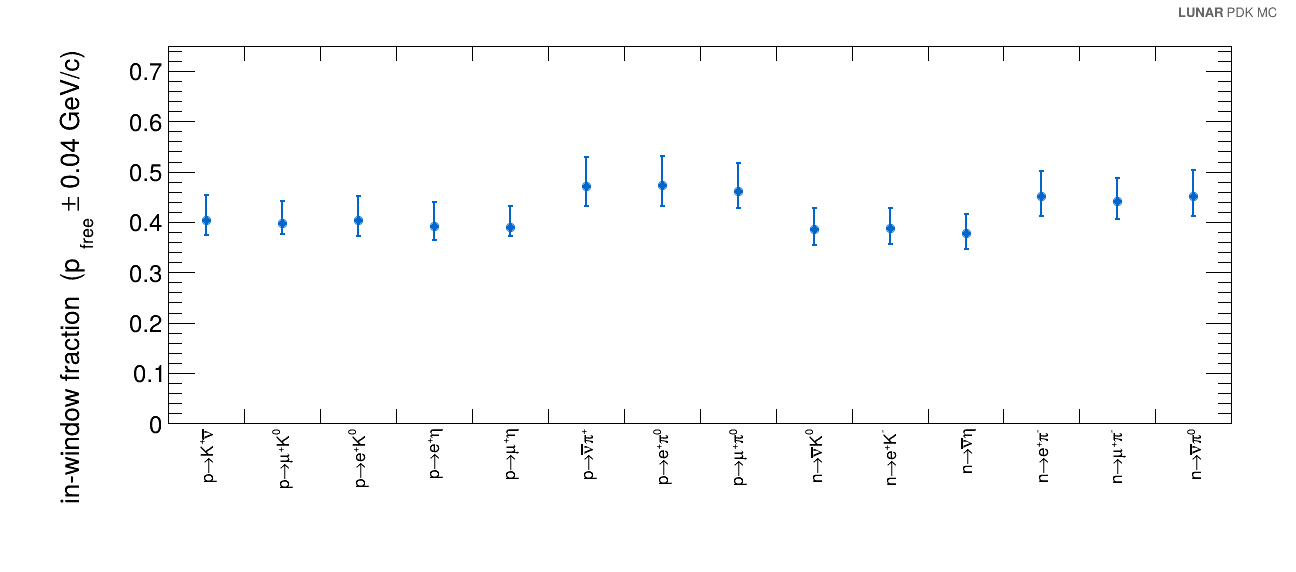}
  \caption{Signal-window containment efficiency $\varepsilon_{\text{win}}$,
  defined in the text, by channel. The marker is the mean over the ten physics
  momentum models and the bar their full spread. Only the relative spread is
  used, as a proxy for the nuclear-model dependence of a spectrum-based
  selection, and it is what sets the nuclear-model band on the event rates of
  Table~\ref{tab:predictions}. Each of the fourteen channels is generated with
  each of the ten momentum models, $5\times10^{4}$ events per combination,
  cascade off.}
  \label{fig:windoweff}
\end{figure*}

Anchoring $\varepsilon_{\text{det}}$ to the DUNE assumption this way rests on a
cascade consistent with DUNE's own: our $K^{+}$ survival ($\sim$99\%,
Table~\ref{tab:fsi}) agrees with the
$\sim$97\% that the GENIE-based DUNE simulation finds for the kaon escaping the
argon nucleus, and our $\pi^{0}$ result, roughly half the pions reworked, is
consistent with the DUNE $p\to e^{+}\pi^{0}$ efficiency being limited by inelastic
intranuclear scattering~\cite{DUNE2020}. The cascade survival
$\varepsilon_{\text{FSI}}$ runs from $0.99$ for the $S=+1$ kaons to $\sim$0.42
for the $K^{-}$ and is the dominant channel-to-channel modulation of the
predicted counts.

Two uncertainty bands accompany each count, and it is worth being explicit about
what they
represent, because neither is a complete systematic. The first is the
\emph{nuclear-model} band. A selection in a liquid-argon detector will not be a
cut on momentum; it will use $\mathrm{d}E/\mathrm{d}x$, track topology, and the
decay chain of the stopping meson. Any such selection is nonetheless tuned on a
simulated signal spectrum, and it is the model dependence of that spectrum we
want to expose. We therefore use a deliberately simple, transparent stand-in: the
\emph{signal-window containment efficiency}
$\varepsilon_{\text{win}}$, defined as the fraction of decays whose hadron-daughter
laboratory momentum falls within $\pm 0.04\GeV/c$ of the free-decay value
$p_{\text{free}}$ for that channel. It is not an absolute reconstruction
efficiency and is not applied as one; only its \emph{relative} spread across the
ten momentum models is used, as a proxy for how much a spectrum-based selection
efficiency would move if the nuclear model were changed. Figure~\ref{fig:windoweff}
shows $\varepsilon_{\text{win}}$ channel by channel; the spread is
$^{+11}_{-8}\%$ about the model average, and we carry it as a relative band on
$\varepsilon$, and hence on $N$. Because the spread enters as a ratio, treating it
as a band on $\varepsilon_{\text{det}}$ or on $N$ is the same statement; we quote
it on $N$ because that is the quantity of interest.

The second is the \emph{FSI-model} band. Since LUNAR provides one cascade
(Sec.~\ref{sec:fsi}), we bound its modeling uncertainty with the formation-zone
scan of Table~\ref{tab:fz}, which is the ingredient that moves
$\varepsilon_{\text{FSI}}$ most. The predictions use the default cascade
(formation zone off) as the central value, and the band extends upward to the
$c\tau_f=1.0\,$fm result. For the pion modes this band is several times wider
than the nuclear-model band, the point at which the FSI treatment, not the
nuclear ground state, becomes the limiting systematic.

Neither band includes the binding prescription (which shifts the central value of
the spectrum, not the rate), the value of $\varepsilon_{\text{det}}$ itself, or
the atmospheric-neutrino background. The counts $N$ are thus model-band estimates
of the detected signal under a fixed, uniform efficiency assumption, not a
complete systematic budget.

\begin{table*}[tbp]
  \centering
  \begin{tabular}{@{}llccc rrr@{}}
    \toprule
    Mode & $N$ & $\tau/B$ limit & $\varepsilon_{\text{det}}$ &
    $\varepsilon_{\text{FSI}}$ & $N_{\text{evt}}$ & $N_{\text{evt}}$ & $N_{\text{evt}}$ \\
         &     & [yr] (90\% C.L.) & & & (FSI on) & (FZ max) & (FSI off) \\
    \midrule
    $p\to K^{+}\bar\nu$  & $p$ & $5.9\times10^{33}$~\cite{SuperK2014}      & 0.30 & 0.99 & $5.4^{+0.7}_{-0.4}$  & $5.4$  & $5.5$ \\
    $p\to \mu^{+}K^{0}$  & $p$ & $3.6\times10^{33}$~\cite{SuperK2022mK0}   & 0.30 & 0.99 & $8.9^{+1.0}_{-0.5}$  & $8.9$  & $9.0$ \\
    $p\to e^{+}K^{0}$    & $p$ & $1.0\times10^{33}$~\cite{SuperK2005susy}  & 0.30 & 0.99 & $32^{+4}_{-2}$       & $32$   & $33$  \\
    $p\to e^{+}\eta$     & $p$ & $1.4\times10^{34}$~\cite{SuperK2024eta}   & 0.30 & 0.57 & $1.3^{+0.2}_{-0.1}$  & $1.5$  & $2.3$ \\
    $p\to \mu^{+}\eta$   & $p$ & $7.3\times10^{33}$~\cite{SuperK2024eta}   & 0.30 & 0.57 & $2.5^{+0.3}_{-0.1}$  & $2.8$  & $4.5$ \\
    $p\to \bar\nu\pi^{+}$& $p$ & $3.9\times10^{32}$~\cite{SuperK2014nupi}  & 0.30 & 0.56 & $47^{+6}_{-4}$       & $69$   & $83$  \\
    $p\to e^{+}\pi^{0}$  & $p$ & $2.4\times10^{34}$~\cite{SuperK2020epi0}  & 0.30 & 0.49 & $0.67^{+0.08}_{-0.06}$ & $1.1$ & $1.4$ \\
    $p\to \mu^{+}\pi^{0}$& $p$ & $1.6\times10^{34}$~\cite{SuperK2020epi0}  & 0.30 & 0.48 & $0.98^{+0.12}_{-0.07}$ & $1.6$ & $2.0$ \\
    \addlinespace
    $n\to \bar\nu K^{0}$ & $n$ & $1.3\times10^{32}$~\cite{SuperK2005susy}  & 0.30 & 0.99 & $302^{+31}_{-25}$    & $302$  & $306$ \\
    $n\to \bar\nu\eta$   & $n$ & $1.6\times10^{32}$~\cite{SuperK2017lmeson}& 0.30 & 0.57 & $142^{+15}_{-12}$    & $155$  & $248$ \\
    $n\to \bar\nu\pi^{0}$& $n$ & $1.1\times10^{33}$~\cite{SuperK2014nupi}  & 0.30 & 0.49 & $18^{+2}_{-2}$       & $29$   & $36$  \\
    $n\to \mu^{+}\pi^{-}$& $n$ & $3.5\times10^{33}$~\cite{SuperK2017lmeson}& 0.30 & 0.55 & $6.2^{+0.6}_{-0.5}$  & $9.3$  & $11$  \\
    $n\to e^{+}\pi^{-}$  & $n$ & $5.3\times10^{33}$~\cite{SuperK2017lmeson}& 0.30 & 0.56 & $4.2^{+0.5}_{-0.4}$  & $6.2$  & $7.5$ \\
    $n\to e^{+}K^{-}$    & $n$ & ---                                      & 0.30 & 0.42 & ---                  & ---    & ---   \\
    \bottomrule
  \end{tabular}
  \caption{Predicted nucleon-decay events in DUNE ($20$~kt,
  $400\ \mathrm{kt\cdot yr}$) with $\tau$ set to each mode's current
  Super-Kamiokande 90\%~C.L.\ limit, so the counts are upper bounds allowed by
  present data. Every mode uses the same factorization
  $\varepsilon=\varepsilon_{\text{det}}\varepsilon_{\text{FSI}}$ with a flat
  $\varepsilon_{\text{det}}=0.30$. The quoted uncertainty on the FSI-on count is
  the nuclear-model band of Fig.~\ref{fig:windoweff}. ``FZ max'' is the upper edge
  of the FSI-model band, obtained with the formation zone at $c\tau_f=1.0\,$fm
  (Table~\ref{tab:fz}); it exceeds the nuclear-model band for every pion mode. The
  last column is the count with the cascade off ($\varepsilon_{\text{FSI}}=1$), so
  the gap to the FSI-on column is what the cascade removes.
  $n\to e^{+}K^{-}$ has no dedicated limit. Limit values follow the PDG
  compilation~\cite{PDG2024}.}
  \label{tab:predictions}
\end{table*}

\begin{figure*}[tbp]
  \centering
  \includegraphics[width=0.92\textwidth]{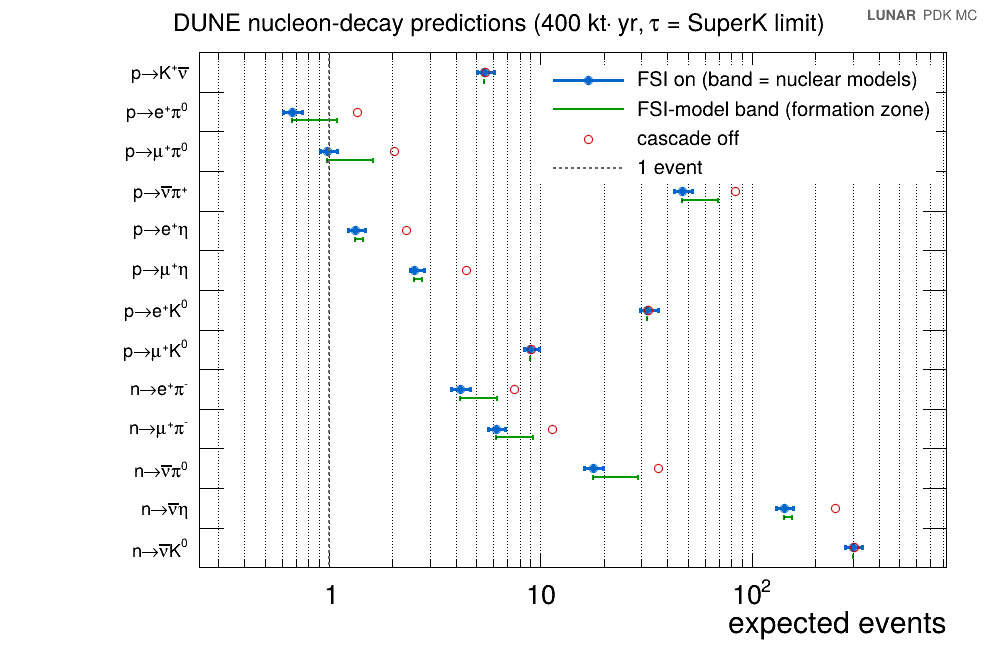}
  \caption{Predicted events per channel in $400\ \mathrm{kt\cdot yr}$ at $\tau$
  equal to the current Super-Kamiokande limit. Filled markers carry the
  nuclear-model band (thick, blue); the thin green bar below each point is the
  FSI-model band, running from the default cascade up to the formation zone at
  $c\tau_f=1.0\,$fm. Open markers are the FSI-unfolded
  counts ($N/\varepsilon_{\text{FSI}}$), so the gap to the filled marker is the
  cascade suppression: substantial for the pions and the $\eta$, negligible for
  the $S=+1$ kaons, whose two bands and two markers coincide. The
  dashed line marks one event.}
  \label{fig:predictions}
\end{figure*}

The results are collected in Table~\ref{tab:predictions} and
Fig.~\ref{fig:predictions}. Even saturating the present limits, DUNE would gather
only a handful of $p\to K^{+}\bar\nu$ events ($\sim$5 in
$400\ \mathrm{kt\cdot yr}$), the golden mode, for which the cascade matters at
the percent level. The channels with the weakest current limits, %
$n\to\bar\nu K^0$, $n\to\bar\nu\eta$, and $p\to\bar\nu\pi^+$, have the most room
for discovery, with tens to a few hundred allowed events. For the pion and $\eta$
modes the decisive systematic is the cascade, which roughly halves the rate, an
effect several times larger than the $^{+11}_{-8}\%$ nuclear-model band, and
whose own modeling uncertainty, bounded here by the formation-zone band, is
itself larger than that band: for $p\to e^{+}\pi^{0}$ the count moves from $0.67$
to $1.1$ across the formation-zone range, against $\pm0.07$ from the nuclear
model. Where the FSI treatment is uncertain at this level, refining the nuclear
ground state buys little. The kaon channels are barely touched by either: with
the $S=+1$ kaons transparent, both $p\to K^{+}\bar\nu$ and the $K^{0}$ modes are
limited by the nuclear model rather than the cascade. Final-state
interactions are thus simultaneously a leading correction and a strongly
mode-dependent one, and they reinforce the strange channels, $p\to K^{+}\bar\nu$
above all, as the cleanest nucleon-decay probes available to a liquid-argon
detector.

A recent GiBUU study by Yan \textit{et al.}~\cite{Yan2026} reaches a
complementary conclusion. Treating $p\to e^{+}\pi^{0}$ in water Cherenkov
detectors with full Boltzmann transport, they find the pion final-state-%
interaction uncertainty to be moderate, while the choice of Fermi-momentum
distribution dominates the systematic, specifically on the atmospheric-neutrino
background rate, a quantity our signal-only generator does not address. The two
pictures are consistent rather than contradictory: both identify the nuclear
initial state and final-state interactions as the controlling nuclear effects,
and which one dominates depends on the target (argon versus water), the channel,
and whether the limiting quantity is the signal efficiency and spectrum or the
background rate. Our argon, signal-side scan across all channels puts the FSI
cascade foremost, since it is what separates the transparent $K^{+}$ from the
heavily reworked pions, whereas their water, background-side study of the single
$e^{+}\pi^{0}$ mode foregrounds the Fermi-momentum distribution. Much of the
contrast is one of metric: their quoted final-state-interaction uncertainty is
the \emph{variation} under FSI-model changes rather than its absolute size, and
that absolute size (${\sim}40\%$ of pions absorbed for production momenta of
$0.4$--$0.5\GeV/c$) in fact agrees with our $\pi^{0}$ result. Barbu \textit{et
al.}~\cite{Barbu2025} survey the same Fermi-motion and FSI effects across argon,
xenon, and water for $p\to\pi^{+}\bar\nu$ and $p\to K^{+}\bar\nu$ and reach the
same qualitative ordering, the kaon far less affected than the pion, an
independent cross-check of the cross sections and mean free paths our cascade
encodes; their analytic full-nucleus attenuation, with isospin-averaged
$K$--$N$ cross sections and a harder argon Fermi momentum ($241$ versus our
$217\MeV/c$), nonetheless leaves the argon $K^{+}$ somewhat less transparent than
our data-driven cascade.

\section{Using the generator}
\label{sec:usage}

LUNAR is intended to be picked up and used quickly. The build has no
dependencies beyond a C++17 compiler, and the supplied pipeline compiles the
generator, produces events, fills a ROOT tree, and draws the standard spectra in
a single command. A minimal session selects a channel, a momentum model, a
binding model, and the cascade:
\begin{quote}\small\ttfamily
./run.sh 100000 1 lfg potential pToKnu
\end{quote}
generates $10^5$ $p\to K^{+}\bar\nu$ events with the local Fermi gas and the
optical potential. The momentum model (\code{--model}), binding model
(\code{--binding}), channel (\code{--channel}), and cascade (\code{--fsi
on|off}) are all chosen at run time; the complete option list and the output
format are given in Appendix~\ref{app:usage}. With the cascade enabled the
generator writes a full post-interaction final state, one row per escaping
particle, with its species, laboratory four-momentum, and a label recording the
fate of the primary meson, so that downstream analyses see the same hadronic
activity a detector would.

This lightness is the point: where the general-purpose generators carry
sophisticated nuclear models but are built around the neutrino-interaction vertex
and are heavy to deploy for a nucleon-decay study, LUNAR does one thing and
exposes every ingredient. Three uses follow naturally. First, signal efficiency: because the
generator reports the surviving meson and the accompanying nucleons and pions, it
provides the per-channel acceptance and the hadronic activity that drive a
liquid-argon selection. Second, systematics: the ten momentum models and three
binding prescriptions span the nuclear-model space, and the formation-zone scan
brackets the cascade, so an analyzer can read off
defensible uncertainty bands, as in the $^{+11}_{-8}\%$ nuclear-model spread and
the wider FSI-model band of Sec.~\ref{sec:predictions}, rather than guess them. Third, fast turnaround and
pedagogy: a full model scan runs in seconds on a laptop, which makes the tool
suitable both for rapid sensitivity estimates and for teaching the nuclear physics
of nucleon decay.

\paragraph*{Code availability.}
LUNAR is released as open-source software at
\url{https://github.com/LUNARPDK/lunarpdk} under a permissive license, together
with the analysis macros, the spectral-function grids, and the scripts that
produce every figure in this paper. The release is intended to let the community
reproduce these
results and adapt the generator to detectors and channels beyond those studied
here.

\section{Possible extensions}
\label{sec:extensions}

The modular, header-only design makes LUNAR straightforward to extend, and
several directions are natural. The nuclear input is the most localized: changing
the density parameters, Fermi momenta, and shell tables retargets the generator
to a different nucleus, so the same machinery could serve the oxygen of
water-Cherenkov detectors, for direct cross-checks against Super-Kamiokande and
Hyper-Kamiokande, or other liquid-argon variants, limited only by the
availability of a tabulated spectral function.

The decay vertex can be broadened beyond the two-body modes treated here.
Three-body and radiative final states, and the inclusive cascade decays of heavier
mesons, would extend the channel list, while dinucleon decay and
neutron--antineutron oscillation~\cite{PhillipsNNbar2016} would let the same
nuclear and cascade machinery address baryon-number violation by two units. Each
amounts to adding an entry to the channel registry and the corresponding
phase-space generator.

The cascade itself admits systematic refinement. An in-medium $K^+$ potential,
additional inelastic channels, and diffractive slopes fitted channel by channel
would sharpen the strange-sector transport; a complete treatment of the secondary
decays of escaping $\pi^0$, $\eta$, $K^0_L$, and hyperons, together with the
de-excitation photons and nucleons emitted as the excited residual nucleus
relaxes (extra activity that a liquid-argon detector can tag for background
rejection, and that must be modeled for realistic efficiency estimates) would
complete the visible final state. On the nuclear side, fitted rather than
representative proton and
neutron Fermi momenta and shell energies would tighten the model band.

Finally, interoperability would broaden the generator's reach. Writing events in
a standardized record such as HepMC3~\cite{BuckleyHepMC2021} or the
neutrino-specific NuHepMC~\cite{NuHepMC2023} would let LUNAR feed detector
simulations directly, and a thin interface to a liquid-argon simulation chain
would close the loop between the nuclear physics modeled here and a full detector
response. None of these requires structural change; each is a local addition
enabled by the present design.

\section{Summary}
\label{sec:summary}

We have presented LUNAR, a compact and openly available Monte Carlo generator for
bound-nucleon decay in argon-40. It samples the decaying nucleon from one of
ten nuclear momentum models, binds it off the mass shell with one of three
removal-energy prescriptions (including the momentum-dependent optical potential
of Ref.~\cite{Juszczak2005}), performs the off-shell two-body decay into the
laboratory frame, and transports the daughter meson out of the nucleus through a
semi-classical intranuclear cascade, writing a complete final state for any of
the standard proton and neutron channels.

Using the generator we have separated the nuclear effects that shape the
observable meson spectrum, finding that the momentum model alone controls its
width while both the momentum and the binding model move its central value, the
former by the larger amount; we have quantified
final-state interactions mode by mode, showing that the nucleus is nearly
transparent to the $S=+1$ kaons but absorbs roughly half of the pions, $\eta$
mesons, and antikaons; and we have translated current Super-Kamiokande limits into
expected DUNE yields across all standard channels, where the cascade emerges as
the dominant systematic for every mode except the strange ones. These
results reinforce $p\to K^{+}\bar\nu$ as the cleanest liquid-argon nucleon-decay
probe and, for the other modes, identify final-state interactions, and the
choice of FSI model rather than the nuclear ground state, as the leading
correction.

The generator has the expected limitations of a fast, dedicated tool. Most of the
analytic momentum models are parametrized rather than fitted; the cascade uses
tabulated cross sections and representative angular slopes rather than a
channel-by-channel fit; and the detector response is not included. None of these
is structural, and the modular design we have emphasized makes each a candidate
for the refinements of Sec.~\ref{sec:extensions}. We release LUNAR to the
community in that spirit, as a transparent, extensible instrument for the nuclear
physics of nucleon decay in liquid argon.

\section*{Acknowledgments}
This work was supported by the Science and Technology Facilities Council (STFC) Lancaster EPP Consolidated Grant 2025-2029:  UKRI2846.

\appendix

\section{Momentum-model parameters}
\label{app:models}

The analytic momentum models of Table~\ref{tab:models} use the following choices.
The short-range-correlation (\code{src}) and toy spectral-function (\code{sf})
models place a default 20\% of nucleons in a $1/p^4$ tail extending to
$k_{\max}=0.65\GeV/c$. The Bodek--Ritchie gas (\code{br})~\cite{BodekRitchie1981}
uses a larger 25\% fraction reaching to $1\GeV/c$, and the correlated Fermi gas
(\code{cfg}) a 20\% contact tail to $2\kf$. The harmonic-oscillator shell model
(\code{hosm}) builds $n(p)$ from oscillator wavefunctions using the $(n,l)$
quantum numbers of the argon shell table, and the Gaussian (\code{gauss}) is a
smooth baseline with $\sigma=\kf/\sqrt5$ and no Fermi surface. The argon nuclear
parameters (the density of Eq.~\eqref{eq:density}, the Fermi momenta
$\kf^{p}=217\MeV/c$ and $\kf^{n}=230\MeV/c$, the isospin fractions, and the proton and
neutron shell levels used by the \code{shell} binding model) follow
Ref.~\cite{Juszczak2005}. The \code{benhar} option instead draws
$(p,E_{\mathrm{rem}})$ jointly from the tabulated argon spectral function of
Refs.~\cite{JLabArgonSF2022,BanerjeeAnkowski2024}, and \code{ankowski} from an
analytic effective spectral function (a $f_{\mathrm{corr}}=0.20$ correlated
$1/p^4$ tail plus a shell mean field built from the argon orbitals); both ignore
the binding parameters above.

\section{Command-line options and output format}
\label{app:usage}

The generator is driven by the executable \code{LunarPDKGenerator} with the
following options: \code{--events} (number of decays), \code{--channel} (mode key
of Table~\ref{tab:channels}), \code{--model} (momentum model of
Table~\ref{tab:models}), \code{--binding} (\code{potential}, \code{constant}, or
\code{shell}), \code{--fsi} (\code{on} or \code{off}), \code{--decay-mesons}
(decay escaping $\pi^0$, $\eta$, and $K^0$), \code{--formation-zone} and
\code{--formation-time} (free-stream produced mesons one formation length before
re-interaction), \code{--seed}, and \code{--sf-file} (override the
spectral-function grid).

The \code{--sf-file} option accepts any tabulated $S(p,E)$ in the NuWro
\code{gridfun2d} plain-text format, so the \code{benhar} model is not restricted
to the argon grids shipped with the code. The file begins with three header
lines (the number of removal-energy points and momentum points, the two axis
minima, and the two axis maxima, all energies and momenta in MeV) followed by
one block per momentum value: the momentum, then the $(E, S)$ pairs at that
momentum, energy index running fastest. Absolute normalization is irrelevant, as
only ratios enter the sampling. Supplying a grid for a different nucleus is
therefore enough to retarget the initial state, provided the corresponding
density and Fermi-momentum parameters are updated as well
(Sec.~\ref{sec:extensions}).

With the cascade on, the output is one row
per escaping particle (\code{event}, \code{pdg}, the laboratory momentum
components \code{px}, \code{py}, \code{pz}, the energy \code{E}, and a fate label
\code{outcome}), preceded by a per-event header recording the nucleon momentum,
removal energy, and primary-meson fate. With the cascade off, the generator
writes the legacy momentum-only table (\code{event}, \code{nucleon\_p},
\code{d1\_p}, \code{d2\_p}, \code{e\_rem}) read by the single-distribution
analysis macros. All momenta and energies are in GeV.


\end{document}